\documentclass[preprint]{ptephy_v1}

\preprintnumber{XXXX-XXXX} 
\usepackage{graphicx}	
\usepackage{amsmath}	
\usepackage{amssymb}	
\usepackage{natbib}
\usepackage{hyperref}
\usepackage{subfigure}
\usepackage{romannum}
\usepackage{cancel}
\usepackage{ulem}

\newcommand{\diff}{{\text{d}}}

\usepackage{color}

\begin{document}

\title{
Steady-State Solutions for a Geometrically Thin Accretion Disk with Magnetically-Driven Winds
}


\author{Mageshwaran Tamilan}
\affil{Department of Space Science and Astronomy, Chungbuk National University, Cheongju 361-763, Korea 
\email{tmageshwaran@chungbuk.ac.kr}}

\author[1,2]{Kimitake Hayasaki}
\affil{Department of Physical Sciences, Aoyama Gakuin University, Sagamihara 252-5258, Japan}

\author[3]{Takeru K. Suzuki}
\affil{School of Arts and Sciences, The University of Tokyo, 3-8-1, Meguro, Tokyo 153-8902, Japan}


\begin{abstract}%
We present steady-state solutions for a one-dimensional, magnetically-driven accretion disk wind model based on magnetohydrodynamic equations. We assume a geometrically thin, gas-pressure-dominated accretion disk, incorporating both magnetic braking and turbulent viscosity introduced by an extended alpha-viscosity prescription. Additionally, the vertical stress parameter is assumed to scale with the disk aspect ratio. We confirm that the derived solutions result in standard disk solutions when the wind is absent. We find that the mass accretion rate decreases as the disk mass falls inward, while the mass loss rate increases with radius. The disk spectrum emitted from the magnetically-driven disk wind can be observed without interference from the wind medium because the wind is significantly optically thin. The spectral luminosity is proportional to $\nu^{1/3}$ in the intermediate, multicolor-blackbody wavebands, in the absence of wind, as predicted by standard disk theory. However, in the presence of wind, it follows a different power-law dependence on frequency over the same range. A deviation from the spectral slope of $1/3$, particularly a negative spectral slope, is a clear indicator of the presence of a magnetically driven wind. We also discuss an observational strategy to test our model with multi-wavelength observations.
\end{abstract}

\subjectindex{E10, E14, E31, E34, E36}

\maketitle

%


%
\section{Introduction} 
\label{sec:intro}
%

Diverse astronomical systems, such as X-ray binaries (XRBs), active galactic nuclei (AGNs), and tidal disruption events (TDEs), are powered by the accretion of matter onto black holes through accretion disks. Historically, hydrodynamic turbulence was thought to provide the effective viscosity, facilitating the redistribution of angular momentum within the disk. This viscous process leads to the inward radial flow of matter toward the black hole, while a small fraction of the mass flows outward. By introducing the $\alpha$-parameter to represent the turbulent viscosity, Shakura \& Sunyaev (1973)~\cite{1973A&A....24..337S} developed a steady-state solution for a geometrically thin, optically thick accretion disk. In the standard disk model, energy conservation dictates that the disk temperature depends on the radius and surface density, given by $T \propto r^{-1/2}\Sigma^{2/3}$, where $r$ is the radius, and $T$ and $\Sigma$ are the midplane temperature and surface density of the disk, respectively. The spectral luminosity exhibits multicolor blackbody emission with a characteristic spectral slope of $\nu^{1/3}$ in the frequency $\nu$, extending from the mid-infrared to the soft X-ray bands \cite{1981ARA&A..19..137P,2002apa..book.....F,2008bhad.book.....K}.

In the presence of a magnetic field in the disk, the differential rotation of electrically conducting fluids around a central object causes a magnetorotational instability (MRI; \cite{Velikhov1959,1961hhs..book.....C,1991ApJ...376..214B,1998RvMP...70....1B}). The resulting magnetohydrodynamic (MHD) turbulence generates Maxwell and Reynolds stresses, which drive mass accretion in the disk. Notably, Balbus \& Papaloizou (1999)~\cite{1999ApJ...521..650B} have shown that the mean flow dynamics of MHD turbulence in a disk obey the $\alpha$ prescription. Moreover, the MHD turbulent stresses also launch vertical outflows, the intensity of which depends on the magnetic field strength in the disk \cite{2009ApJ...691L..49S,2014ApJ...784..121S,2013ApJ...765..149C,2019ApJ...872..149L}. These MHD-driven outflows remove mass, angular momentum, and energy from the disk, thereby affecting its structure and emission.

Suzuki et al. (2016)\cite{2016A&A...596A..74S} constructed a one-dimensional (1D), time-dependent, geometrically thin accretion disk model with magnetically driven winds, specifically for protoplanetary disks. In their model, the MHD turbulence stresses were parameterized in both the radial and vertical directions using an extended $\alpha$ prescription. Subsequently, Tabone et al. (2022)\cite{2022MNRAS.512.2290T} derived a self-similar solution for a similar 1D, time-dependent, geometrically thin accretion disk with magnetically driven winds, also in the context of protoplanetary disks. They assumed that the disk temperature follows a proportionality of $r^{-1/2}$, and that the vertical stress $\alpha$-parameter is proportional to the disk's aspect ratio. In contrast, Mageshwaran et al. (2024)~\cite{2024ApJ...975...94T} studied a similar 1D, time-dependent, geometrically thin disk with magnetically driven winds in the context of TDEs, and obtained the time evolution of an initial Gaussian disk with the inner boundary condition of a zero-torque at the innermost stable circular orbit (ISCO) radius. They found that the mass accretion rates evolve more steeply than $t^{-19/16}$ \cite{1990ApJ...351...38C} in late times, with the power-law index of time saturating at late times. In contrast, the mass accretion rate becomes significantly steeper than $t^{-2}$ in the presence of magnetic braking, i.e., when the vertical stress parameter $\alpha$ has a non-zero value. This case can explain the light curve dips as $t^{-2.54}$ observed in the TDE candidate AT2019qiz \cite{2020MNRAS.499..482N}.

While time-dependent studies of magnetically-driven disk winds, such as those applied to TDEs, have provided significant insights into accretion dynamics, steady-state solutions are crucial for understanding systems where accretion occurs over extended timescales with a steady mass supply from the outer region, such as AGNs and XRBs. These systems exhibit relatively stable disk structures, allowing for the application of steady-state models to explore long-term wind behavior and mass loss mechanisms. Steady-state solutions are key for interpreting the observed spectra and outflows from AGNs and XRBs, where the winds, driven by magnetic fields, play a pivotal role in regulating the angular momentum transport and accretion rates. Blandford \& Payne (1982) \cite{1982MNRAS.199..883B} and Ferreira \& Pelletier (1995) \cite{1995A&A...295..807F} demonstrated the significance of magneto-centrifugally driven winds in these scenarios, while Fukumura et al. (2010) \cite{2010ApJ...715..636F} and Miller et al. (2016) \citep{2016ApJ...821L...9M} investigated the observational signatures of such winds in AGNs or XRBs.

This motivate our focus on constructing steady-state models for the magnetically-driven disk winds, which are not only vital for understanding the long-term evolution of these accretion systems but also for providing predictive models that can be directly compared with observational data across various wavelengths. In this paper, we analytically derive a steady-state solution for a 1D, time-dependent, geometrically thin disk with a magnetically-driven wind, using an extended $\alpha$ parameterization. We assume that the vertical stress $\alpha$-parameter is proportional to the disk aspect ratio. In Section \ref{sec:model}, we present the basic equations of our model. In Section \ref{sec:steadystate}, we outline the derivation of the steady-state solution and discuss the resulting radial disk structures, disk spectra, and the optical depth of the wind. The limitations and observational implications of our model are discussed in Section~\ref{sec:dis}. Finally, Section~\ref{sec:con} is devoted to our conclusions.

%
\section{Basic Equations} 
\label{sec:model}
%

We consider a one-dimensional, time-dependent, axisymmetric disk with a magnetically-driven wind, which has been constructed by \cite{2016A&A...596A..74S,2024ApJ...975...94T}. As the disk model, we apply a vertically integrated, geometrically thin, viscous accretion disk with a Keplerian rotation $\Omega=\sqrt{GM/r^3}$ where $M$ is the black hole mass and $G$ is the gravitational constant. Following their studies, the vertically integrated mass conservation equation is given by
\begin{equation}
\label{eq:sigt}
\frac{\partial \Sigma}{\partial t} 
+ 
\frac{1}{r}\frac{\partial}{\partial r}(r \Sigma v_r) + \dot{\Sigma}_{\rm w}  = 0,
\end{equation}
where $\dot{\Sigma}_{\rm w}$ is the vertical mass flux of the wind, $\Sigma = 2 H \rho$ is the surface density in the disk {with the disk's mass density $\rho$}, $H$ is the disk scale height and $v_r$ is the radial velocity. For a Keplerian rotation angular velocity, the angular momentum conservation given by equation (\ref{cylmcons1}) results in
\begin{equation}\label{eq:rsvr}
r \Sigma v_r = -\frac{2}{r\Omega} 
\left[\frac{\partial}{\partial r} 
\left( 
\bar{\alpha}_{r\phi} r^2 \Sigma c_s^2 
\right) 
+ 
\frac{\bar{\alpha}_{z\phi}}{2} 
\frac{r^2 \Sigma c_s^2}{H}
\right],
\end{equation} 
where the sound speed of the disk is given by
\begin{equation}\label{eq:cs}
c_s 
= 
\sqrt{\frac{k_{\rm B} T }{ \mu m_p}},
\end{equation} 
$T$ is the mid-plane temperature of the disk, $k_{\rm B}$ is the Boltzmann constant, $m_{p}$ is the proton mass, $\mu$ is the mean molecular weight taken to be ionized solar mean molecular weight of $0.65$, and $\bar{\alpha}_{r\phi}$ and $\bar{\alpha}_{z\phi}$ are introduced as parameters due to the MHD turbulence and disk winds (the so-called extended $\alpha$ prescription), which are given by equations (\ref{eq:alrphi0}) and (\ref{eq:alzphi1}) respectively.

The evolution equation of surface density is given by substituting equation~(\ref{eq:rsvr}) into equation~(\ref{eq:sigt}) as 
\begin{equation}\label{eq:sdevo}
\frac{\partial \Sigma}{\partial t} 
- 
\frac{2}{r}\frac{\partial}{\partial r}
\left[
\frac{1}{r\Omega}
\left\{
\frac{\partial}{\partial r} 
\left(\bar{\alpha}_{r\phi} r^2 \Sigma c_{s}^2 \right) + \bar{\alpha}_{z\phi} r^2 \rho c_{s}^2 
\right\}
\right] 
+ 
\dot\Sigma_{\rm w}=0.
\end{equation}

The MHD energy equation for a geometrically thin disk with Keplerian rotational velocity is given by 
\begin{equation}\label{eq:ene0}
\dot{\Sigma}_{\rm w} \left[E_{\rm w} + \frac{r^2 \Omega^2}{2} \right] + {Q}_{\rm rad} = Q_+
\end{equation}
where 
$E_{\rm w}$ denotes the wind energy, which is given by
\begin{equation}\label{eq:ew0}
E_{\rm w} = \frac{1}{2} v^2 + \Phi + \frac{\gamma c_s^2}{\gamma -1} + \frac{B_{\phi}^2 + B_r^2}{4\pi \rho} - \frac{B_z}{4\pi\rho v_z} \left(v_{\phi}B_{\phi} + v_r B_r\right),
\end{equation}
$Q_{\rm rad}$ is the radiative cooling flux, which is given by
\begin{equation}\label{eq:qrad}
Q_{\rm rad} = \frac{64\sigma T^4}{3\kappa_{\rm es}\Sigma}
\end{equation}
with $\sigma$ as the Stefan-Boltzmann constant and $\kappa_{\rm es}=0.34~{\rm cm^{2}~g^{-1}}$ as the Thomson scattering opacity\footnote{
See Appendix~\ref{app:opacity} for the more general opacity formula.},
and 
$Q_+$ represents the heating flux due to turbulent viscosity and magnetic braking as
\begin{equation}\label{eq:qvisf}
Q_{+} = 
\frac{3}{2}
\bar{\alpha}_{r\phi} 
\Omega\,
\Sigma\,c_{s}^2 
+ 
\frac{1}{2} \bar{\alpha}_{z\phi} r \Omega^2\Sigma c_{s}.
\end{equation}

The wind energy, denoted by $E_{\rm w}$, represents the total wind energy per unit mass. It consists of the specific kinetic, gravitational, thermal, and magnetic energies, as well as the Poynting flux normalized by the mass flux. The essential condition for a disk to have wind is $E_{\rm w} \geq 0$, which ensures that the wind material can escape to infinity. In our calculation, we obtain a particular wind solution by considering $E_{\rm w} = 0$ \cite{2016A&A...596A..74S,2024ApJ...975...94T}. This assumption implies that the magnetic energy, including the contribution from the Poynting flux, plays a critical role in driving the wind against the gravitational potential. Thus, the magnetic fields act as an energy intermediaries, converting the gravitational energy released by accretion into both the kinetic energy of the wind and the radiative cooling energy of the disk.
Equation~(\ref{eq:ene0}) is then reduced to 
\begin{equation}
\label{eq:ene1}
\dot{\Sigma}_{\rm w} 
\frac{r^2 \Omega^2}{2} 
+
Q_{\rm rad} 
= Q_{+}.
\end{equation}
Since equations (\ref{eq:sdevo}) and (\ref{eq:ene1}), which are a set of basic equations for our model, have three variables ($\Sigma, T, \dot{\Sigma}_{\rm w}$), the two equations are impossible to solve as they are. Therefore, we introduce an additional equation to make a closure of those equations as
\begin{equation}\label{eq:qradf}
Q_{\rm rad} 
= 
\epsilon_{\rm rad}
\, 
Q_+,
\end{equation}
where $\epsilon_{\rm rad}$ means the fraction of the heating flux of the disk that is converted into radiative cooling flux, and is treated as a parameter in the range of $0<\epsilon_{\rm rad}\le1$. Substituting equation (\ref{eq:qradf}) into equation (\ref{eq:ene1}) yields the vertical mass flux as
\begin{eqnarray}
\label{eq:sigdotw}
\dot{\Sigma}_{\rm w} 
&&
=
\frac{2}{r^2\Omega^2}(Q_+-Q_{\rm rad})
\nonumber \\
&&
= 
(1-\epsilon_{\rm rad})\left[3\bar{\alpha}_{r\phi} \frac{c_s^2\Sigma}{r^2\Omega} + \bar{\alpha}_{z\phi}\frac{c_s \Sigma}{r}\right].
\end{eqnarray} 
It is evident that the discrepancy between the heating and radiative cooling fluxes accounts for the wind flux. Equation~(\ref{eq:sdevo}) is rewritten using equation~(\ref{eq:sigdotw}) as
\begin{equation}\label{eq:sdevo2}
\frac{\partial \Sigma}{\partial t} - \frac{2}{r}\frac{\partial}{\partial r}\left[\frac{1}{r\Omega}\left\{\frac{\partial}{\partial r} \left(r^2 \Sigma \bar{\alpha}_{r\phi} c_{s}^2 \right) + \frac{\bar{\alpha}_{z\phi}}{2}r^2 \Omega \Sigma c_{s} \right\}\right] + (1-\epsilon_{\rm rad})\left[3\bar{\alpha}_{r\phi} \frac{c_s^2\Sigma}{r^2\Omega} + \bar{\alpha}_{z\phi}\frac{c_s \Sigma}{r}\right]=0.
\end{equation}
The equation to determine the sound speed is given using equations (\ref{eq:cs}), (\ref{eq:qrad}), and (\ref{eq:qradf}) by
\begin{equation}\label{eq:cstazphi}
c_s^7 = \frac{9\kappa_{\rm es}}{128 \sigma} \left(\frac{k_B}{\mu m_p}\right)^{4}\bar{\alpha}_{r\phi} \epsilon_{\rm rad} r \Omega^2 \Sigma^2 \left[\frac{c_s}{r\Omega} + \frac{\bar{\alpha}_{z\phi}}{3\bar{\alpha}_{r\phi}} \right].
\end{equation}
The hydrostatic balance of the disk is given by
\begin{equation}\label{eq:thindisk} 
c_{s} = \Omega{H}.
\end{equation}
Equations (\ref{eq:sdevo2}), (\ref{eq:cstazphi}), and (\ref{eq:thindisk}) are a set of basic equations and need to be solved simultaneously to study how the accretion disk with magnetically driven wind evolves.

Mageshwaran et al. (2024)~\cite{2024ApJ...975...94T} performed the numerical simulations for constant $\bar{\alpha}_{z\phi}$ in the context of TDEs, where they assumed an initial Gaussian disk with total disk mass equal to the half of stellar mass and surface density at the inner and the outer edge of the disk to be zero. They found that non-zero $\bar{\alpha}_{z\phi}$ results in a rapid decline in the surface density compared to the case of $\bar{\alpha}_{z\phi} = 0$ due to increased mass accretion rate and the vertical wind mass flux. The mass accretion and wind rates' power-law index do not saturate to specific values and decline below $< -2$ at late times. 
This implies that no self-similar solution exists for a non-zero value of $\bar{\alpha}_{z\phi}$.
However, the analytical solutions can exist even for the $\bar{\alpha}_{z\phi}\neq0$ case, if the magnetic braking parameter is assumed to be
\begin{equation}\label{eq:azphi}
\bar{\alpha}_{z\phi} = \alpha_{\rm c} \frac{H}{r},
\end{equation}
where $\alpha_{\rm c}$ is a constant free parameter, indicating that the $\bar{\alpha}_{z\phi}$ evolves with radius and time. In fact, Tabone et al. (2022) \cite{2022MNRAS.512.2290T} introduced the parameters $\alpha_{\rm SS}$ and $\alpha_{\rm DW}$ to represent the radial and vertical stresses, respectively, and derived self-similar solutions. A comparison with our $\alpha$ parameterization shows that $\bar{\alpha}_{r\phi}=(3/2)\alpha_{\rm SS}$ and $\bar{\alpha}_{z\phi}=(3/2)\alpha_{\rm DW}(H/r)$, leading to $\alpha_{\rm c} = (3/2) \alpha_{\rm DW}$, where equation~(\ref{eq:azphi}) was applied.
Substituting equations~(\ref{eq:thindisk}) and (\ref{eq:azphi}) into equation (\ref{eq:sigdotw}), we obtain the vertical mass flux as
\begin{equation}\label{eq:sigdotw2}
\dot{\Sigma}_{\rm w} = 3\left(1-\epsilon_{\rm rad}\right)\bar{\alpha}_{r\phi}\left(1+\frac{\psi}{3}\right)\frac{c_{s}^2 \Sigma}{r^2\Omega},
\end{equation}
where we introduced the fractional parameter $\psi$ as   
\begin{equation}\label{eq:psi}
\psi = \frac{\alpha_{\rm c}}{\bar{\alpha}_{r\phi}}.
\end{equation}

The vertical mass flux can be also expressed by
\begin{equation}\label{eq:sigdotw-tb}
\dot{\Sigma}_{\rm w} 
=
\frac{1}{2}
\frac{\alpha_{\rm c}}{\lambda-1}
\frac{c_{s}^2 \Sigma}{r^2\Omega},
\end{equation}
where $\lambda$ is the magnetic lever arm parameter \cite{1982MNRAS.199..883B,2022MNRAS.512.2290T}. 
Equating equation (\ref{eq:sigdotw2}) with equation (\ref{eq:sigdotw-tb}) yields the relation among $\lambda$, $\epsilon_{\rm rad}$, and $\psi$ as

\begin{eqnarray}\label{eq:lambda}
\lambda =1 + \frac{\psi}{2(1-\epsilon_{\rm rad})(3+\psi)}.
\end{eqnarray}
There are four cases depending on whether the wind is blowing and whether magnetic braking is working. For each case, the possible values or ranges of the three parameters are presented in Table~\ref{tbl:1}.

%
%
\begin{table}[!h]
\caption{
Possible values or ranges of the three parameters to determine whether the wind is blowing with or without the influence of magnetic braking.
}
\label{tbl:1}
\centering
\begin{tabular}{cccccc}
\hline
\hline
 & $\epsilon_{\rm rad}$ & $\psi$ & $\lambda$ & {\rm Wind} & {\rm Magnetic Braking} \\
\hline
\hline
Case 1 & $1$ & $0$ & $1$ & {\rm No} & {\rm Off} \\
Case 2 & $1$ & $>0$ & $\infty$ & {\rm No} & {\rm On} \\
Case 3 & $<1$ & $0$ & $1$ & {\rm Yes} & {\rm Off} \\
Case 4 & $<1$ & $>0$ & $>1$ & {\rm Yes} & {\rm On} \\
\hline
\hline
\end{tabular}
\end{table}

The vertical mass flux increases with a decrease in $\epsilon_{\rm rad}$ and with an increase in $\psi$. 
Substituting equations~(\ref{eq:thindisk}), (\ref{eq:azphi}), and (\ref{eq:psi}) into equation~(\ref{eq:qvisf}) yields
\begin{equation}\label{eq:qvisf2}
Q_{+} = 
\frac{3}{2}
\bar{\alpha}_{r\phi} 
\left(
1
+
\frac{\psi}{3}
\right)
\Omega\,
\Sigma\,c_{s}^2.
\end{equation}
Substituting equations~(\ref{eq:azphi}), (\ref{eq:thindisk}), and (\ref{eq:sigdotw2}) into equation~(\ref{eq:sdevo2}) yields 
\begin{equation}\label{eq:sdevo3}
\frac{\partial \Sigma}{\partial t} - \frac{2}{r}\frac{\partial}{\partial r}
\left[
\frac{1}{r\Omega}
\left\{
\frac{\partial}{\partial r} 
\left(\bar{\alpha}_{r\phi}  r^2 \Sigma c_{s}^2 \right)
+ 
\frac{1}{2} \psi\bar{\alpha}_{r\phi} r \Sigma c_{s}^2 
\right\}
\right] 
+
3\left(1-\epsilon_{\rm rad}\right)\bar{\alpha}_{r\phi}\left(1+\frac{\psi}{3}\right)\frac{c_{s}^2 \Sigma}{r^2\Omega}=0.
\end{equation}
Using equations~(\ref{eq:thindisk}), (\ref{eq:azphi}), and (\ref{eq:psi}), equation~(\ref{eq:cstazphi}) is also simplified to be
\begin{equation}\label{eq:cstazphi1}
c_s^6 = \frac{9}{128}\frac{\kappa_{\rm es}}{ \sigma} \left(\frac{k_{\rm B}}{\mu m_p}\right)^{4} 
\epsilon_{\rm rad} 
\bar{\alpha}_{r\phi} 
\left(1+ \frac{\psi}{3} \right)
\Omega \Sigma^2.
\end{equation}
Equations (\ref{eq:sdevo3}) and (\ref{eq:cstazphi1}) form the foundation of our disk-wind model. By solving these equations, we obtain the steady-state solution, which is described in the subsequent sections.

%
\section{Steady state solutions}
\label{sec:steadystate}
%

In this section, we {derive} a steady-state solution with a zero torque inner boundary condition. {The mass accretion rate is defined by
\begin{eqnarray}\label{eq:maccrate}
 \dot{M} = -2\pi r\Sigma v_r.
\end{eqnarray}
Since the partial derivative with time is zero, i.e, $\partial\Sigma/\partial t=0$ in the steady state, 
equation (\ref{eq:sigt}) is rewritten by
\begin{equation}\label{eq:dmdr}
\frac{{\rm d}\dot{M}}{{\rm d} r} =  2\pi r\dot{\Sigma}_{\rm w} = 6 \pi \left(1-\epsilon_{\rm rad}\right)\bar{\alpha}_{r\phi}\left(1+\frac{\psi}{3}\right)\frac{c_{s}^2 \Sigma}{r\Omega}.
\end{equation}
where we used equations (\ref{eq:sigdotw2}) and (\ref{eq:maccrate}) for the derivation.
Using equations (\ref{eq:rsvr}), (\ref{eq:azphi}), and (\ref{eq:psi}), equation~(\ref{eq:maccrate}) is rewritten as
\begin{equation}\label{eq:maccrate2} 
\dot{M} =  \frac{4\pi \bar{\alpha}_{r\phi}}{\Omega} r^{-1-\frac{\psi}{2}} \frac{{\rm d}}{{\rm d }r}\left[r^{2+\frac{\psi}{2}} c_s^2 \Sigma\right].
\end{equation}
Combining equation~(\ref{eq:dmdr}) with equation~(\ref{eq:maccrate2}) yields the following second-order differential equation on $\dot{M}$ as
\begin{equation}
\frac{{\rm d}}{{\rm d}r}\left[r^{\frac{3+\psi}{2}}\frac{{\rm d} \dot{M}}{{\rm d}r}\right] = \frac{3}{2} (1-\epsilon_{\rm rad}) \left[1+\frac{\psi}{3}\right] \dot{M} r^{\frac{-1+\psi}{2}}
\nonumber 
\end{equation} 
for the case of a Keplerian disk. The solution is then given by
\begin{equation}\label{eq:mdotstgeneral}
\dot{M} = C_1 r^{\zeta_1}  + C_2 r^{-\zeta_2},
\end{equation}
where $C_1$ and $C_2$ are constant of integrations, and $\zeta_1$ and $\zeta_2$ are given by
\begin{eqnarray}
\zeta_1 
&&
= \frac{\delta-1 -\psi}{4}
\label{eq:zeta1}
\\
\zeta_2
&&
=
\frac{\delta+1 +\psi}{4}
\label{eq:zeta2}
\end{eqnarray}
with
\begin{equation}\label{eq:delta}
\delta = \sqrt{(1+\psi)^2+8(3+\psi)(1-\epsilon_{\rm rad})}.
\end{equation}
We note that $\zeta_1\ge0$, $\zeta_2>0$, and $\delta>0$. Applying the two boundary conditions that $\dot{M}=\dot{M}_{\rm out}$ at the outer edge radius of the disk $r_{\rm out}$ for a given constant $\dot{M}_{\rm out}$ and the zero torque at the inner edge radius of the disk $r_{\rm in}$, meaning that $\diff \dot{M}/\diff r = 0$  at $r = r_{\rm in}$ (see equation \ref{eq:dmdr})
, we can determine the integral constants as
\begin{eqnarray}
C_1
&&
=
\frac{
\dot{M}_{\rm out}
}
{
r_{\rm out}^{\zeta_1}
}
\biggr[
1+
\left(
\frac{\zeta_1}{\zeta_2}
\right)
\left(\frac{r_{\rm in}}{r_{\rm out}}\right)^{\delta/2}
\biggr]^{-1}, 
\label{eq:c1}
\\
C_2
&&
=
\frac{\zeta_1}{\zeta_2}
\frac{
\dot{M}_{\rm out}
}
{
r_{\rm out}^{\zeta_1}
}
r_{\rm in}^{\delta/2}
\biggr[
1+
\left(
\frac{\zeta_1}{\zeta_2}
\right)
\left(\frac{r_{\rm in}}{r_{\rm out}}\right)^{\delta/2}
\biggr]^{-1}. 
\label{eq:c2}
\end{eqnarray} 
Substituting them into equation~(\ref{eq:mdotstgeneral}), we obtain the mass accretion rate as
\begin{equation}
\dot{M} =  
\dot{M}_{\rm out}
\left(\frac{r}{r_{\rm out}}\right)^{\zeta_1}
\left[
1
+
\left(
\frac{\zeta_1}{\zeta_2}
\right) 
\left(\frac{r_{\rm in}}{r}\right)^{\delta/2}
\right]
\biggr[
1+
\left(
\frac{\zeta_1}{\zeta_2}
\right)
\left(\frac{r_{\rm in}}{r_{\rm out}}\right)^{\delta/2}
\biggr]^{-1}.
\label{eq:mdot}
\end{equation}
While the mass accretion rate goes to $\dot{M}_{\rm out}$ in the absence of wind ($\epsilon_{\rm rad} = 1$), the mass accretion rate increases with radius according to equation~(\ref{eq:mdot}) in the presence of wind ($0 < \epsilon_{\rm rad} < 1$).
}

Integrating equation (\ref{eq:dmdr}) over the disk as
\begin{equation}
\int_{r_{\rm in}}^{r} \frac{\diff \dot{M}}{\diff r}\,\diff r = \int_{r_{\rm in}}^{r} 2 \pi r \dot{\Sigma}_{\rm w} \, \diff r
\nonumber
\end{equation}
yields
\begin{equation}\label{eq:mdotcons}
\dot{M}(r)-\dot{M}_{\rm in} = \dot{M}_{\rm w}(r),
\end{equation}
where $\dot{M}_{\rm in}$ is the mass accretion rate estimated at $r_{\rm in}$ and the mass loss rate is defined by
\begin{eqnarray} 
\dot{M}_{\rm w}(r) = \int_{r_{\rm in}}^{r} 2 \pi r \dot{\Sigma}_{\rm w} \, \diff r.
\label{eq:mdotw0}
\end{eqnarray} 
Substituting equation~(\ref{eq:mdot}) into equation~(\ref{eq:mdotcons}) gives
\begin{equation}\label{eq:mdotw}
\dot{M}_{\rm w}(r) 
=
\dot{M}_{\rm out} \biggr[
1+
\left(
\frac{\zeta_1}{\zeta_2}
\right)
\left(\frac{r_{\rm in}}{r_{\rm out}}\right)^{\delta/2}
\biggr]^{-1} \left(\frac{r}{r_{\rm out}}\right)^{\zeta_1}
\Biggr[
1+\frac{\zeta_1}{\zeta_2}\left(\frac{r_{\rm in}}{r}\right)^{\delta/2} - \frac{\delta}{2\zeta_2}\left(\frac{r_{\rm in}}{r}\right)^{\zeta_1} 
\Biggr].
\end{equation}

The radial derivative of equation~(\ref{eq:mdotstgeneral}) is given by 
\begin{equation}\label{eq:dmdr2}
\frac{d\dot{M}}{dr} = C_1\zeta_1 r^{\zeta_1-1}  - C_2\zeta_2 r^{-\zeta_2-1}=C_1\zeta_1r^{\zeta_1-1}
\left[
1
-
\left(
\frac{r_{\rm in}}{r}
\right)^{\delta/2}
\right].
\end{equation}
Equating equation~(\ref{eq:dmdr}) with equation~(\ref{eq:dmdr2}) using equations~(\ref{eq:c1}) and (\ref{eq:c2}) gives
\begin{equation}\label{eq:cssigma2}
c_s^2\Sigma 
= 
\frac{\dot{M}_{\rm out}\Omega_{\rm out}}{4 \pi\bar{\alpha}_{r\phi}\,\zeta_2 }
\left[
1+
\frac{\zeta_1}{\zeta_2}
\left(
\frac{r_{\rm in}}{r_{\rm out}}
\right)^{\delta/2}
\right]^{-1} 
 \left(\frac{r}{r_{\rm out}}\right)^{\zeta_1-3/2} 
 \left[1-\left(\frac{r_{\rm in}}{r}\right)^{\frac{\delta}{2}}\right],
\end{equation}
where $\Omega_{\rm out} = \sqrt{G M / r_{\rm out}^3}$ and, for the derivation, we used $\zeta_1\zeta_2=(3+\psi)(1-\epsilon_{\rm rad})/2$, which are obtained from equations~(\ref{eq:zeta1})-(\ref{eq:delta}). Substituting equation (\ref{eq:cssigma2}) into equation~(\ref{eq:qvisf2}) yields the total heating flux in the disk as
\begin{eqnarray}
Q_{+} 
=
\frac{3+\psi}{8 \pi \zeta_2}\dot{M}_{\rm out}\Omega_{\rm out}^2   
\left[
1+
\frac{\zeta_1}{\zeta_2}
\left(
\frac{r_{\rm in}}{r_{\rm out}}
\right)^{\delta/2}
\right]^{-1} 
 \left(\frac{r}{r_{\rm out}}\right)^{\zeta_1-3} 
 \left[1-\left(\frac{r_{\rm in}}{r}\right)^{\frac{\delta}{2}}\right],
 \label{eq:qplus}
\end{eqnarray}
No wind solution with ($\epsilon_{\rm rad},\,\psi)=(1,0)$ results in $\zeta_1=0$, $\zeta_2 = 1/2$, and $\delta=1$, which reduces equation~(\ref{eq:qplus}) to
\begin{equation}
Q_{+} = \frac{3}{4 \pi} \frac{G M \dot{M}_{\rm out}}{r^3} \left[1 - \left(\frac{r_{\rm in}}{r}\right)^{\frac{1}{2}} \right].
\nonumber
\end{equation}
This equation agrees with the standard thin disk solution \cite{1973A&A....24..337S,2002apa..book.....F,2008bhad.book.....K}. By substituting equation~(\ref{eq:qplus}) into equation~(\ref{eq:qradf}), $Q_{\rm rad}$ is rewritten as
\begin{equation}\label{eq:qrad2}
Q_{\rm rad} = 
\epsilon_{\rm rad}
\frac{3+\psi}{8 \pi \zeta_2}\dot{M}_{\rm out}\Omega_{\rm out}^2   
\left[
1+
\frac{\zeta_1}{\zeta_2}
\left(
\frac{r_{\rm in}}{r_{\rm out}}
\right)^{\delta/2}
\right]^{-1} 
 \left(\frac{r}{r_{\rm out}}\right)^{\zeta_1-3} 
 \left[1-\left(\frac{r_{\rm in}}{r}\right)^{\frac{\delta}{2}}\right].
\end{equation}
The relation between the surface density and the radiative cooling flux is given by using equations (\ref{eq:cs}) and (\ref{eq:qrad}) as 
\begin{eqnarray}
\Sigma 
= 
\left(
\frac{
\mu m_{\rm p}
}{
k_{\rm B}
}
\right)^{4/5}
\left(
\frac{64\sigma}{3\kappa_{\rm es}}\right)^{1/5}
\left(
\frac{
(c_s^2\Sigma)^{4}
}{
Q_{\rm rad}
}
\right)^{1/5}.
\nonumber
\end{eqnarray} 
Combining this equation with equations~(\ref{eq:cssigma2}) and (\ref{eq:qrad2}) gives the surface density as
\begin{multline}
\Sigma = \left(\frac{2\sigma}{3\pi^3\kappa_{\rm es}}\right)^{1/5} \left(\frac{\mu m_p}{k_{\rm B}}\right)^{4/5} 
\frac{\dot{M}_{\rm out}^{3/5}\Omega_{\rm out}^{2/5}}{(\epsilon_{\rm rad}\bar{\alpha}_{r\phi}^4 (3+\psi) \zeta_2^3)^{1/5}}
\left[
1+
\frac{\zeta_1}{\zeta_2}
\left(
\frac{r_{\rm in}}{r_{\rm out}}
\right)^{\delta/2}
\right]^{-3/5} \\
 \left(\frac{r}{r_{\rm out}}\right)^{3(\zeta_1-1)/5} 
 \left[1-\left(\frac{r_{\rm in}}{r}\right)^{\delta/2}\right]^{3/5}.
\label{eq:sigmast}
\end{multline}
%
Because $T = (\mu m_p/k_{\rm B}) (c_s^2\Sigma/\Sigma)$, we obtain the disk mid-plane temperature as
\begin{multline}
T
=
\frac{1}{4}
\left(\frac{3}{2\pi^2} \frac{\kappa_{\rm es}}{\sigma} \frac{\mu m_p}{k_{\rm B}}\right)^{1/5} 
\left(
\frac{\epsilon_{\rm rad} (3+\psi) \dot{M}_{\rm out}^{2}\Omega_{\rm out}^{3}}{\bar{\alpha}_{r\phi} \zeta_2^2}
\right)^{1/5}
\left[
1+
\frac{\zeta_1}{\zeta_2}
\left(
\frac{r_{\rm in}}{r_{\rm out}}
\right)^{\delta/2}
\right]^{-2/5} \\
 \left(\frac{r}{r_{\rm out}}\right)^{2\zeta_1/5-9/10}
 \left[1-\left(\frac{r_{\rm in}}{r}\right)^{\delta/2}\right]^{2/5}.
\label{eq:tmidst}
\end{multline}
No wind solution with ($\epsilon_{\rm rad},\,\psi)=(1,0)$ results in $\zeta_1=0$, $\zeta_2=1/2$, and $\delta=1$, which reduces equations~(\ref{eq:sigmast}) and (\ref{eq:tmidst}) to
\begin{eqnarray}
\Sigma 
&&
=
\left(\frac{16\sigma}{9\pi^3\kappa_{\rm es}}\right)^{1/5} 
\left(\frac{\mu m_p}{k_{\rm B}}\right)^{4/5} 
\frac{\dot{M}_{\rm out}^{3/5}\Omega_{\rm out}^{2/5} }{\bar{\alpha}_{r\phi}^{4/5}} 
 \left(\frac{r}{r_{\rm out}}\right)^{-3/5} 
 \left[1-\left(\frac{r_{\rm in}}{r}\right)^{1/2}\right]^{3/5},
\nonumber \\
T
&&
=
\frac{1}{4}
\left(\frac{18}{\pi^2} \frac{\kappa_{\rm es}}{\sigma} \frac{\mu m_p}{k_{\rm B}}\right)^{1/5} 
\frac{ \dot{M}_{\rm out}^{2/5}\Omega_{\rm out}^{3/5}}{\bar{\alpha}_{r\phi}^{1/5}}
 \left(\frac{r}{r_{\rm out}}\right)^{-9/10} 
 \left[1-\left(\frac{r_{\rm in}}{r}\right)^{1/2}\right]^{2/5},
 \nonumber 
\end{eqnarray}
respectively. For given $M$, $\dot{M}_{\rm out}$, $r_{\rm in}$, and $r_{\rm out}$, these solutions correspond to the standard disk solutions of the electron-scattering-opacity dominated region \cite{1973A&A....24..337S,2002apa..book.....F,2008bhad.book.....K}.

In Appendix \ref{app:opacity}, we derive the solutions by adopting a more general form of the opacity for equation~(\ref{eq:qrad}), including the Rosseland mean opacities approximately expressed by Kramers' law. We also briefly discuss how these solutions differ from those obtained with the Thomson scattering opacity in this section.


%
\subsection{Choice of Physical Parameters}
\label{sec:par}
%

Our model requires seven parameters ($\bar{\alpha}_{r\phi}$, $\epsilon_{\rm rad}$, $\psi$, $M$, $\dot{M}_{\rm out}$, $r_{\rm in}$, $r_{\rm out}$) to represent a specific solution. 
We adopt $\bar{\alpha}_{r\phi}=0.1$ and $\psi=0$ or $\psi=10$ throughout this paper. For the $\epsilon_{\rm rad}$ parameter, we choose three different values from the set $(0.1,\,0.5,\,0.9,\,1.0)$, depending on the specific plots. We take $\dot{M}_{\rm out} = L_{\rm Edd}/c^2$, where $c$ is the speed of light, $L_{\rm Edd}=4\pi GMc/\kappa_{\rm es}$ is the Eddington luminosity and
\begin{equation}
\frac{L_{\rm Edd}}{c^2}
= 1.64\times 10^{24}~{\rm g~s^{-1}} 
 \left(
 \frac{M}{10^7 M_{\odot}} 
 \right)
\end{equation}

The disk inner radius is taken to be the ISCO radius for a non-rotating black hole, 
$r_{\rm in} = 6 r_{\rm g}$, where 
\begin{eqnarray}
r_{\rm g} 
&=& 
\frac{G M}{c^2} 
\nonumber \\
&=& 
1.5 \times 10^{12}~{\rm cm}~\left(\frac{M}{10^7 M_{\odot}} \right)
\end{eqnarray}
is the gravitational radius.

We apply the steady-state solutions to accretion disks in AGNs and XRBs. In AGNs, the size of the accretion disk has been  observationally studied using interband time delays \cite{2022MNRAS.511.3005J,2022ApJ...929...19G}. For the 19 AGN samples, the disk size is  estimated to be $0.1-1$ light-days, which corresponds to $1.74\times 10^3-1.74 \times 10^4~ r_{\rm g}$,  in mass range $10^{6.19} - 10^{8.66} M_{\odot}$. On the other hand, in XRBs with the binary semi-major axis $a_{\rm b}$ with mass ratio $q = M_{\star} / M$, where $M_{\star}$ is the mass of the companion star. Here, the binary orbital period $P_{\rm orb}=2\pi\sqrt{a_{\rm b}^3/(GM)}$, ranging from hour to days \cite{2012ApJ...756...32B,2014RAA....14.1367C}. Using these physical quantities, we obtain the Roche-lobe radius as
\begin{eqnarray}
r_{\rm L} 
&=& 
a_{\rm b} f(q)
\nonumber \\
&\simeq&
3.6 \times 10^{11}~{\rm cm}~
\left(\frac{f(q=0.1)}{0.58}\right)
\left(\frac{M + M_{\star}}{10 M_{\odot}}\right)^{1/3} \left(\frac{P_{\rm orb}}{1 ~{\rm day}}\right)^{2/3},
\end{eqnarray} 
where $f(q)=0.49 q^{-2/3}/[ 0.6 q^{-2/3} + \ln (1+q^{-1/3})]$ \cite{1983ApJ...268..368E} and $f(q=0.1)\simeq0.58$. For $M = 10 M_{\odot}$, $M_{\star} = M_{\odot}$ and $P_{\rm orb} = 1$ day, the Roche lobe radius is $r_{\rm L} \sim 2.5\times 10^5~r_{\rm g}$. In summary, we consider two examples for black hole mass and disk outer radius throughout this paper: one is $M=10^7 M_{\odot}$ and $r_{\rm out}=10^4 r_{\rm g}$, which corresponds to the AGN disk case. The other model is $M=10M_{\odot}$ and $r_{\rm out}=10^5 r_{\rm g}$, which corresponds to the XRB disk case.

%
\subsection{Mass accretion and loss rates}
\label{sec:mdotmw}
%

\begin{figure}
\centering
\subfigure[]{\includegraphics[scale = 0.6]{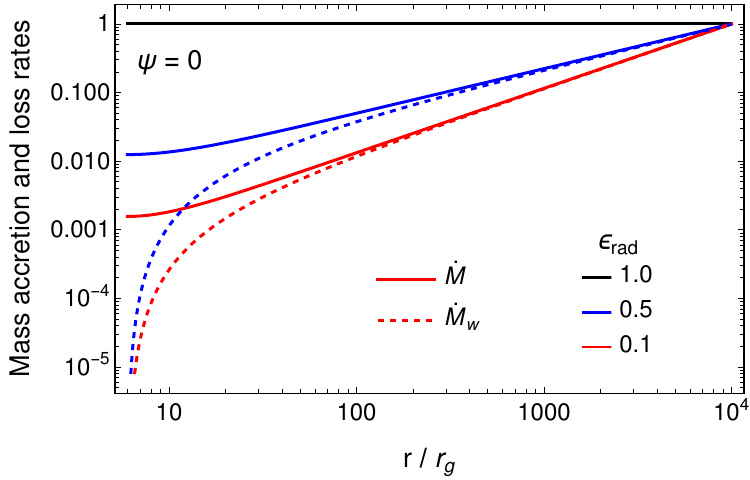}}
\subfigure[]{\includegraphics[scale = 0.6]{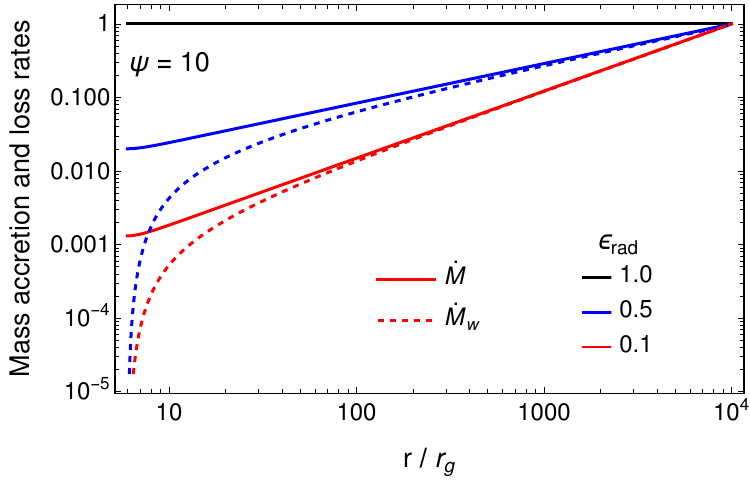}}
\caption{\label{fig:mdotst} 
Radial profiles of the mass accretion and mass loss rates, which are normalized by $L_{\rm Edd} / c^2$. The solid lines represent the mass accretion rate, while the dashed lines show the mass loss rate. The mass accretion and loss rates are calculated for a given boundary condition of $\dot{M}_{\rm out} = L_{\rm Edd}/c^2$ at $r_{\rm out} = 10^4 r_{\rm g}$. Panel (a) shows the case of $\psi=0$, i.e., no magnetic braking, while panel (b) illustrates the case of $\psi=10$ with magnetic braking. In both panels, the black solid line denotes $\dot{M}_{\rm out}$, indicating the case of no wind ($\epsilon_{\rm rad} = 1$). The different colors denote the different values of $\epsilon_{\rm rad}$. 
}
\end{figure}

Figure~\ref{fig:mdotst} shows the radial profiles of mass accretion and loss rates with $r_{\rm out} = 10^4 r_{\rm g}$, and $\dot{M}_{\rm out} = L_{\rm Edd}/c^2$, for different values of $\epsilon_{\rm rad}$ and $\psi$. Equations~(\ref{eq:mdot}) and (\ref{eq:mdotw}) demonstrate that mass accretion and loss rates depend on the black hole mass. However, since the mass accretion and loss rates are normalized by the Eddington accretion rate, and the radius is scaled by the gravitational radius $r_{\rm g}$, the normalized mass accretion and loss rates shown in Figure~\ref{fig:mdotst} are identical regardless of the black hole mass, provided that the same boundary conditions are satisfied. In the absence of the wind, the mass accretion rate remains constant. However, the mass accretion rate decreases as $\epsilon_{\rm rad}$ decreases, and increases with $\psi$ for all values of $\epsilon_{\rm rad}$ beyond a few ISCO radii. Near the ISCO radii, the balance between mass accretion and loss rates determines whether $\dot{M}$ increases with $\psi$.
The mass loss rate increases with radius, and follows the same radial profile as the mass accretion rate at large radii, since $\dot{M}_{\rm w}(r)\sim\dot{M}(r)$ due to $\dot{M}(r)\gg\dot{M}_{\rm in}$ at $r\gg{r_{\rm in}}$ in equation~(\ref{eq:mdotcons}).

%
%

%
\subsection{Disk spectra}
\label{sec:ds}
\begin{figure}
\centering
\subfigure[]{\includegraphics[scale = 0.63]{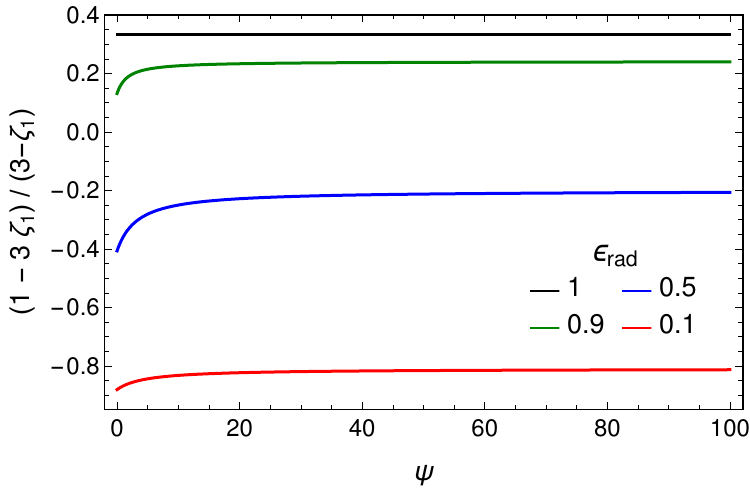}}
\subfigure[]{\includegraphics[scale = 0.63]{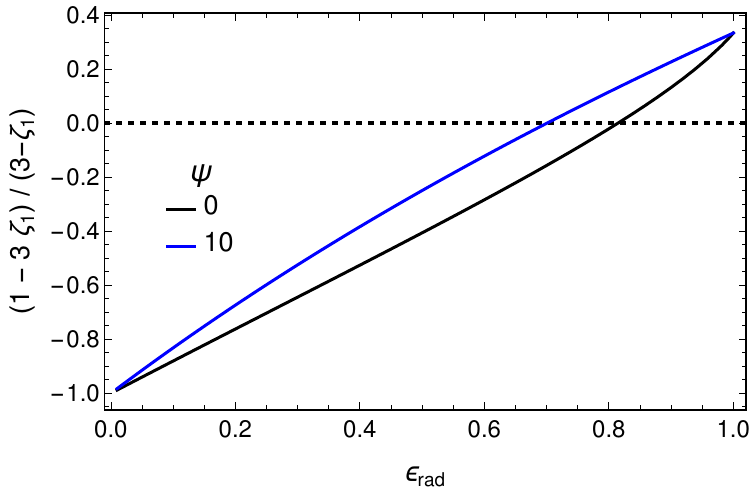}}
\caption{\label{fig:ple} 
Dependence of the power-law index of the flux density, $(1-3\zeta_1)/(3-\zeta_1)$ (equation \ref{eq:spnu}), on $\epsilon_{\rm rad}$ and $\psi$. Panel (a) illustrates the dependence of the power-law index on $\psi$ for various values of $\epsilon_{\rm rad}$. Panel (b) depicts the dependence of the power-law index on $\epsilon_{\rm rad}$ for different values of $\psi$.
}
\end{figure}

%
Since the disk surface temperature is given by $2\sigma T_{\rm eff}^4=Q_{\rm rad}$, equation~(\ref{eq:qrad2}) provides 
\begin{eqnarray}
T_{\rm eff} 
&&
= 
\epsilon_{\rm rad}^{1/4}
\left[
\frac{3+\psi}{16 \pi \sigma \zeta_2}\dot{M}_{\rm out}\Omega_{\rm out}^2   
\right]^{1/4}
\left[
1+
\frac{\zeta_1}{\zeta_2}
\left(
\frac{r_{\rm in}}{r_{\rm out}}
\right)^{\delta/2}
\right]^{-1/4} 
\left(\frac{r}{r_{\rm out}}\right)^{(\zeta_1-3)/4} 
\left[1-\left(\frac{r_{\rm in}}{r}\right)^{\frac{\delta}{2}}\right]^{1/4}
 \nonumber \\
 &&
\approx 
\epsilon_{\rm rad}^{1/4}
\left[
\frac{3+\psi}{16 \pi \sigma \zeta_2}\dot{M}_{\rm out}\Omega_{\rm out}^2   
\right]^{1/4}
\left[
1+
\frac{\zeta_1}{\zeta_2}
\left(
\frac{r_{\rm in}}{r_{\rm out}}
\right)^{\delta/2}
\right]^{-1/4} 
\left(\frac{r}{r_{\rm out}}\right)^{(\zeta_1-3)/4}
\label{eq:teff}
\end{eqnarray}
for $r\gg r_{\rm in}$.
The observed flux of the disk is calculated by integrating the specific intensity, which is the Planck function $B_\nu$, over the surface of the disk as
\begin{eqnarray}
S_\nu
=
\int
B_\nu
\,
d\mathcal{O}
=
4\pi
\frac{h}{c^2}
\frac{\cos{i}}{D^2}
\nu^3
\int_{r_{\rm in}}^{r_{\rm out}}
\,
\frac{r}{e^{h\nu/k_{\rm B}T_{\rm eff}}-1}
dr,
\label{eq:snu}
\end{eqnarray}
where $h$ is the Planck constant, $i$ is the inclination angle of the disk, $D$ is the distance from the earth, and $\mathcal{O}$ is the solid angle subtended by the disk in the observer's sky. The observed flux of the disk becomes 
\begin{equation}\label{eq:spnu}
S_\nu
\propto
\nu^{(1-3\zeta_1)/(3-\zeta_1)}
\end{equation}
for the range of $(k_{\rm B}T_{\rm in}/h)(r_{\rm in}/r_{\rm out})^{(3-\zeta_1)/4}\ll\nu\ll k_{\rm B}T_{\rm in}/h$ \cite{2008bhad.book.....K}. Figure \ref{fig:ple} shows $\psi$-dependence of the power-law index of the flux density, $(1-3\zeta_1)/(3-\zeta_1)$, for different values of $\epsilon_{\rm rad}$. In the absence of the wind (i.e., $\epsilon_{\rm rad} = 1$), the power law index goes to $1/3$, which corresponds to the standard disk case \cite{1981ARA&A..19..137P}. In contrast, in the presence of the wind (i.e., $\epsilon_{\rm rad} < 1$), the power-law index increases with $\epsilon_{\rm rad}$ and $\psi$. This behavior arises because the effective temperature increases with $\epsilon_{\rm rad}$ and $\psi$, driven by changes in the mass accretion rate that affect the viscous heating flux. The increase in the effective temperature results in a higher number of blackbody photons at higher frequencies, shifting the peak frequency (where the flux density shows a significant rise on the high-frequency side) to higher values. As a result, the power-law index increases. The power-law index becomes negative if the following condition is satisfied:
\begin{equation} 
\epsilon_{\rm rad} < \frac{22+6\psi}{27+9\psi}. 
\end{equation} 
This inequality indicates that $\epsilon_{\rm rad} < 22/27$ for $\psi = 0$, while $\epsilon_{\rm rad} < 82/117$ for $\psi = 10$. For $\psi \gg 1$, the power-law index becomes negative when $\epsilon_{\rm rad} < 2/3$.

%
%

%
\begin{figure}
\centering
\subfigure[]{\includegraphics[scale = 0.63]{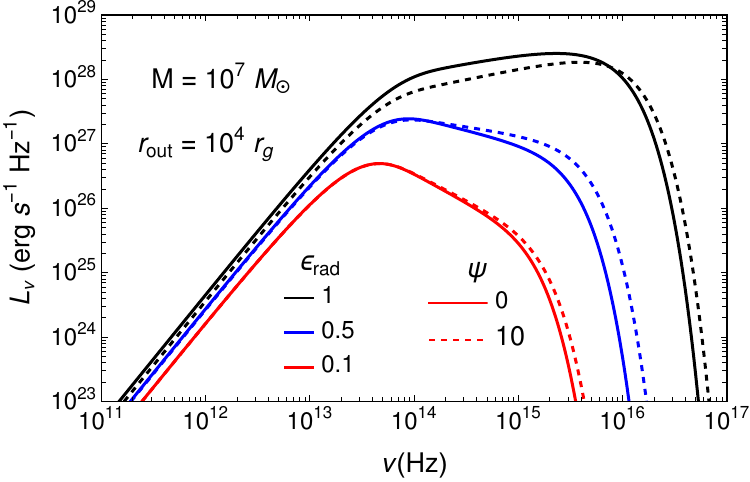}}~~~
\subfigure[]{\includegraphics[scale = 0.63]{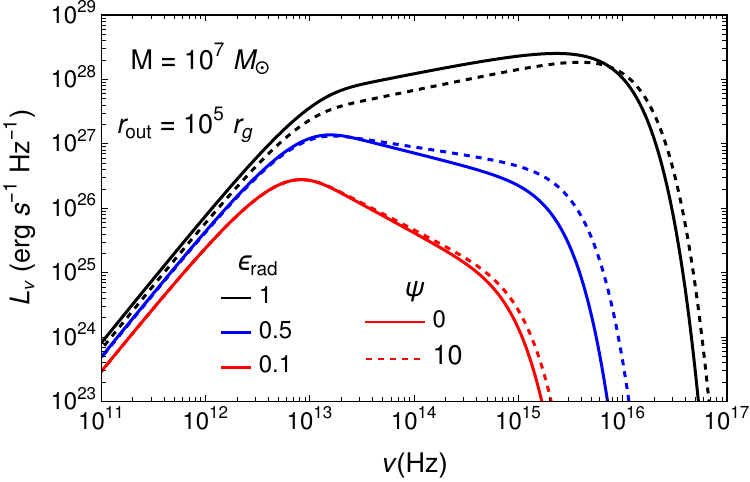}}
\subfigure[]{\includegraphics[scale = 0.63]{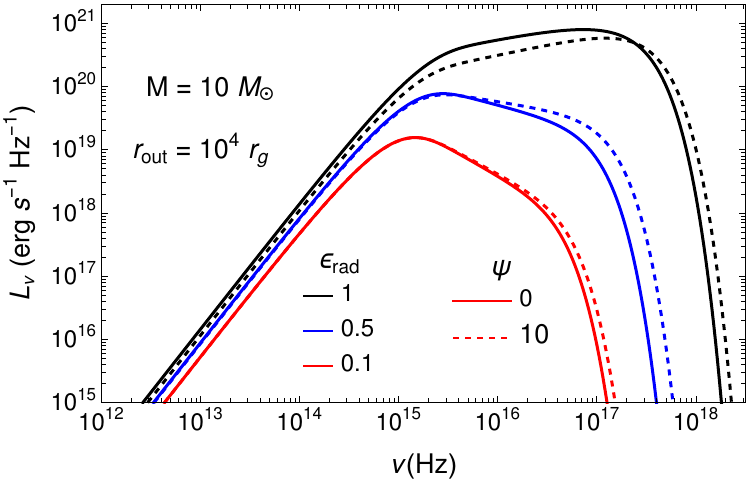}}~~~
\subfigure[]{\includegraphics[scale = 0.63]{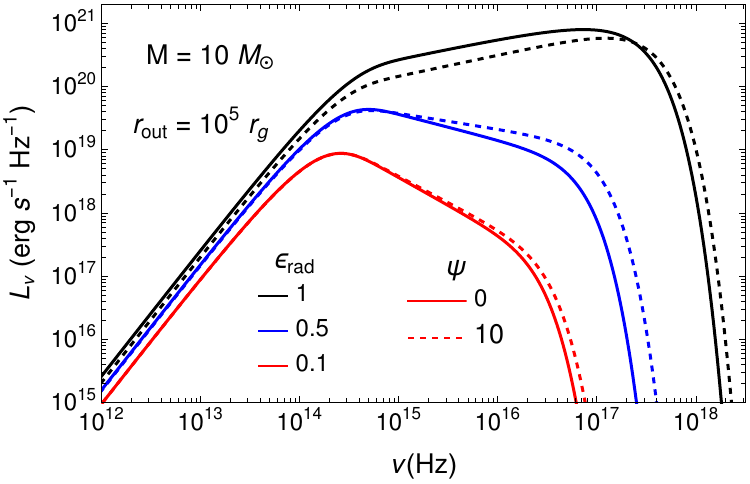}}
\caption{\label{fig:sl} 
Spectral energy distribution of the accretion disk accompanied by wind mass loss. While the different color shows the different value of $\epsilon_{\rm rad}$, the different line type represents the different value of $\psi$. The cases of $\psi =0$ and $\psi=10$ correspond to the absence and presence of magnetic braking, respectively. $\dot{M}_{\rm out} = L_{\rm Edd}/c^2$, $r_{\rm in}=6\,r_{\rm g}$, and $\bar{\alpha}_{r\phi}=0.1$ are adopted. Panels (a) and (b) are for AGN with $M = 10^7 M_{\odot}$ and $r_{\rm out} = 10^4 r_{\rm g}$ and $10^5 r_{\rm g}$, resepctively. Panels (c) and (d) are for XRB with $M = 10 M_{\odot}$ and $r_{\rm out} = 10^4 r_{\rm g}$ and $10^5 r_{\rm g}$, resepctively. 
}
\end{figure}
%

The spectral luminosity is expressed using equation~(\ref{eq:snu}) by $L_{\nu} = 4 \pi D^2 S_\nu$. Figure \ref{fig:sl} illustrates the frequency dependence of $L_\nu$ for both AGN and XRB disks, considering two different values of the outer boundary radius. While the different color denotes the different value of $\epsilon_{\rm rad}$, the different line type represents the different value of $\psi$. It is found from the figure that the disk emission is stronger as $\epsilon_{\rm rad}$ is closer to 1. The luminosity decreases significantly when $\epsilon_{\rm rad}$ decreases from 1 to 0.1, primarily due to the substantial reduction in the mass accretion rate (see Figure \ref{fig:mdotst}), leading to a significant decline in the effective temperature and, consequently, the disk luminosity. The figure also shows an increase in X-ray emission with $\psi$, resulting in a harder spectrum for $\psi=10$ compared to $\psi = 0$. This is because of the increase in the effective temperature with $\psi$, which increases the UV to soft X-ray photons. At low frequencies, the spectral luminosity follows $\nu^2$ due to the Rayleigh-Jeans law, while at high frequencies it decays exponentially in accordance with the Wien law. As predicted in Figure~\ref{fig:ple}, the spectral luminosity increases with frequency for $\epsilon_{\rm rad}=1$ but it decreases with frequency for $\epsilon_{\rm rad}=0.5$ or $0.1$ in the range between the Rayleigh-Jeans and Wien regimes, i.e., in the multicolor blackbody emission regime.

%
\subsection{Physical Quantities of Disk Winds}
\label{sec:dw}
%

Using equations (\ref{eq:ene1}) and (\ref{eq:qradf}), the kinetic energy flux carried by the wind is $Q_{\rm kin}= \dot{\Sigma}_{\rm w} r^2 \Omega^2 / 2 = (1-\epsilon_{\rm rad})Q_{+}$, and using equation (\ref{eq:qplus}), the wind energy flux is given by
\begin{equation}
Q_{\rm kin} = 
\frac{(1-\epsilon_{\rm rad})(3+\psi)}{8 \pi \zeta_2}\dot{M}_{\rm out}\Omega_{\rm out}^2   
\left[
1+
\frac{\zeta_1}{\zeta_2}
\left(
\frac{r_{\rm in}}{r_{\rm out}}
\right)^{\delta/2}
\right]^{-1} 
 \left(\frac{r}{r_{\rm out}}\right)^{\zeta_1-3} 
 \left[1-\left(\frac{r_{\rm in}}{r}\right)^{\frac{\delta}{2}}\right].
\end{equation}  
The kinetic energy carried away by the wind per unit time is denoted by
\begin{eqnarray}\label{eq:Ekin}
\dot{E}_{\rm kin} 
&=& 
\int_{r_{\rm in}}^{r_{\rm out}} Q_{\rm kin} 2 \pi r \, dr 
\nonumber\\
&=&
\frac{(1-\epsilon_{\rm rad})(3+\psi)}{4\zeta_2 (\zeta_1-1)}\dot{M}_{\rm out}\Omega_{\rm out}^2 r_{\rm out}^2 \left[
1+\frac{\zeta_1}{\zeta_2}\left(\frac{r_{\rm in}}{r_{\rm out}}
\right)^{\delta/2}\right]^{-1} 
\nonumber\\
&&
\left[1 + \frac{\zeta_1-1}{\zeta_2+1}\left(\frac{r_{\rm in}}{r_{\rm out}}\right)^{\delta/2}-\frac{\delta}{2(\zeta_2+1)}\left(\frac{r_{\rm in}}{r_{\rm out}}\right)^{\zeta_1-1}\right].
\end{eqnarray}
This is equivalent to the kinetic luminosity, given by $\dot{E}_{\rm kin} = \dot{M}_{\rm w}(r_{\rm out}) v_{\rm w}^2 / 2$, where $v_{\rm w}$ is the average wind velocity and $\dot{M}_{\rm w}(r_{\rm out})$ represents the total mass loss rate, integrated over the entire region of the disk. By using equation (\ref{eq:mdotw}), the total mass loss rate is expressed as
\begin{equation}\label{eq:mwtot}
\dot{M}_{\rm w}(r_{\rm out}) =
\dot{M}_{\rm out} 
\biggr[
1+
\left(
\frac{\zeta_1}{\zeta_2}
\right)
\left(\frac{r_{\rm in}}{r_{\rm out}}\right)^{\delta/2}
\biggr]^{-1} \Biggr[
1+\frac{\zeta_1}{\zeta_2}\left(\frac{r_{\rm in}}{r_{\rm out}}\right)^{\delta/2} - \frac{\delta}{2\zeta_2}\left(\frac{r_{\rm in}}{r_{\rm out}}\right)^{\zeta_1} 
\Biggr],
\end{equation} 
where note that $\dot{M}_{\rm w}(r_{\rm out})$ is not exactly equal to $\dot{M}_{\rm out}$ although $\dot{M}_{\rm w}(r)\sim\dot{M}(r)$ at the large radii as shown in Figure~\ref{fig:mdotst}.
Using equation~(\ref{eq:mwtot}) along with equation (\ref{eq:Ekin}), we get the average wind velocity as
\begin{eqnarray}
v_{\rm w} &=& \sqrt{\frac{2\dot{E}_{\rm kin}}{\dot{M}_{\rm w}(r_{\rm out})}} \nonumber\\
&=& \Omega_{\rm out} r_{\rm out} \left(\frac{(1-\epsilon_{\rm rad})(3+\psi)}{2\zeta_2(\zeta_1-1)}\right)^{1/2}\left[1 + \frac{\zeta_1-1}{\zeta_2+1}\left(\frac{r_{\rm in}}{r_{\rm out}}\right)^{\delta/2}-\frac{\delta}{2(\zeta_2+1)}\left(\frac{r_{\rm in}}{r_{\rm out}}\right)^{\zeta_1-1}\right]^{1/2}  \nonumber \\
&& \left[1+\frac{\zeta_1}{\zeta_2}\left(\frac{r_{\rm in}}{r_{\rm out}}\right)^{\delta/2} - \frac{\delta}{2\zeta_2}\left(\frac{r_{\rm in}}{r_{\rm out}}\right)^{\zeta_1} \right]^{-1/2} 
\label{eq:vw}
\end{eqnarray}
for $0<\epsilon_{\rm rad}< 1$.

\begin{figure}
\centering
\subfigure[]{\includegraphics[scale = 0.6]{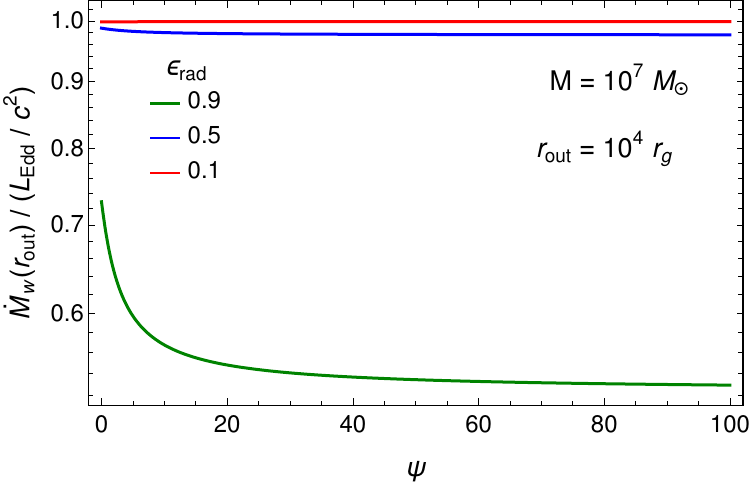}}
\subfigure[]{\includegraphics[scale = 0.65]{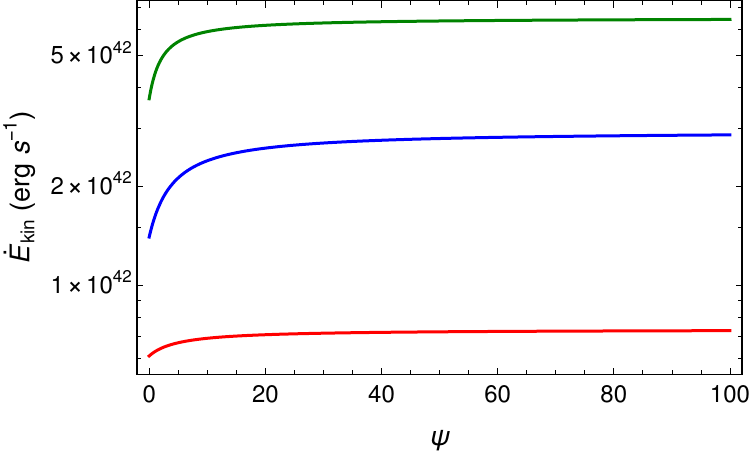}}
\subfigure[]{\includegraphics[scale = 0.6]{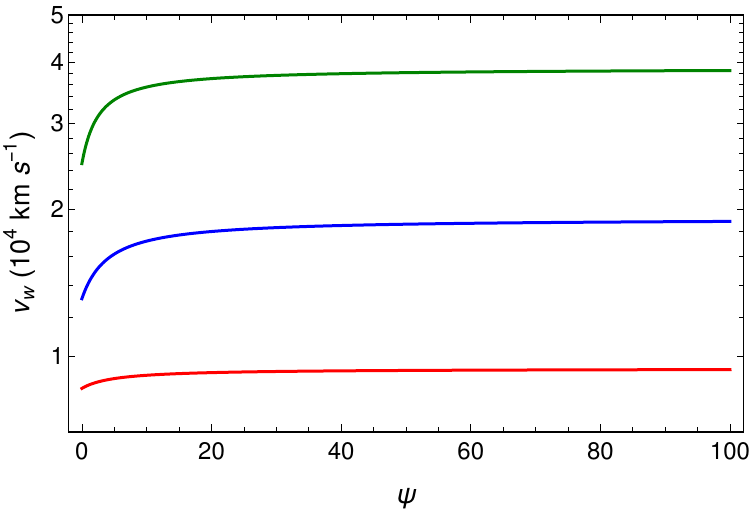}}
\subfigure[]{\includegraphics[scale = 0.65]{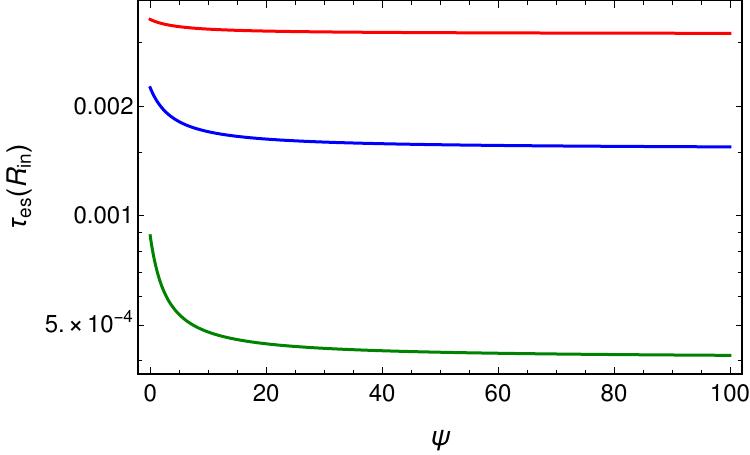}}
\caption{
Dependence of physical quantities of the magnetically-driven disk wind on $\psi$ for three values of $\epsilon_{\rm rad}$. Panels (a), (b), (c), and (d) show the $\psi$-dependence of the wind mass loss rate (see equation \ref{eq:mwtot}), the kinetic energy rate (see equation~\ref{eq:Ekin}), the velocity (see equation \ref{eq:vw}), and the electron scattering optical depth, respectively. The following disk parameters are adopted: $M = 10^7 M_{\odot}$, $\dot{M}_{\rm out} = L_{\rm Edd}/c^2$, $r_{\rm in}=6r_{\rm g}$, and $r_{\rm out} = 10^4 r_{\rm g}$. The wind optical depth, $\tau_{\rm es}$, is estimated using equation (\ref{eq:taues}) at $R_{\rm in}$, which is assumed to be equal to the disk outer radius $r_{\rm out}$.
}
\label{fig:diskwind} 
\end{figure}

\begin{figure}
\centering
\subfigure[]{\includegraphics[scale = 0.42]{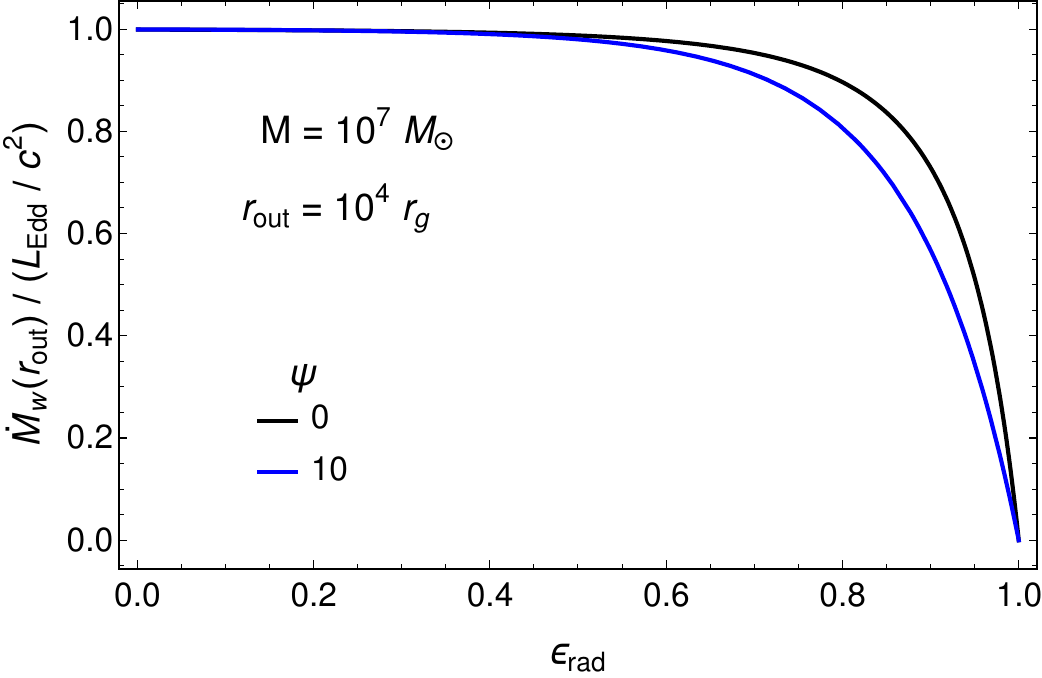}}
\subfigure[]{\includegraphics[scale = 0.45]{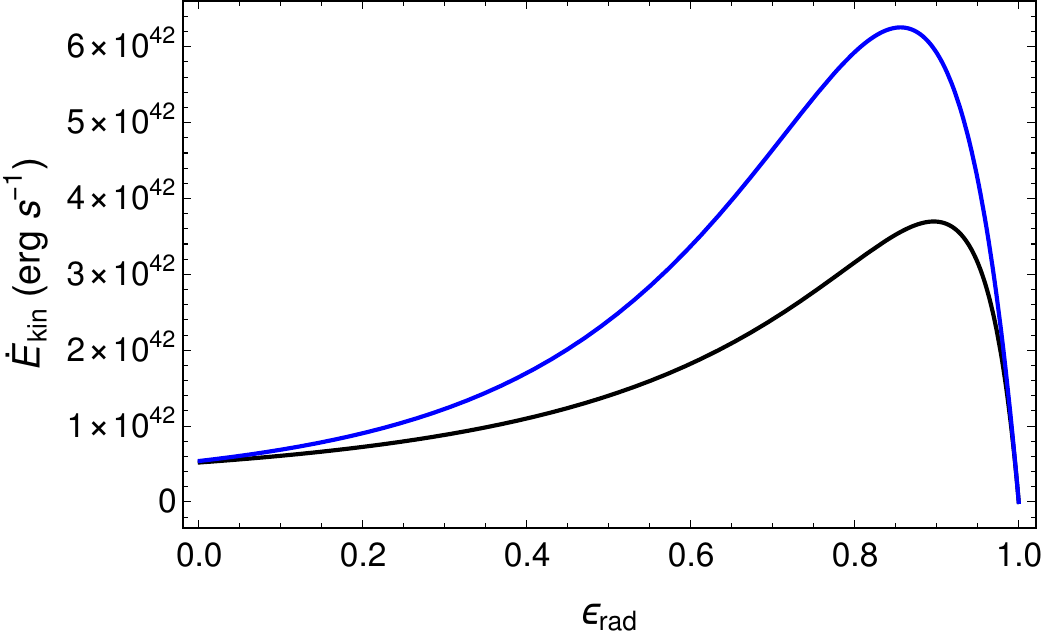}}
\subfigure[]{\includegraphics[scale = 0.43]{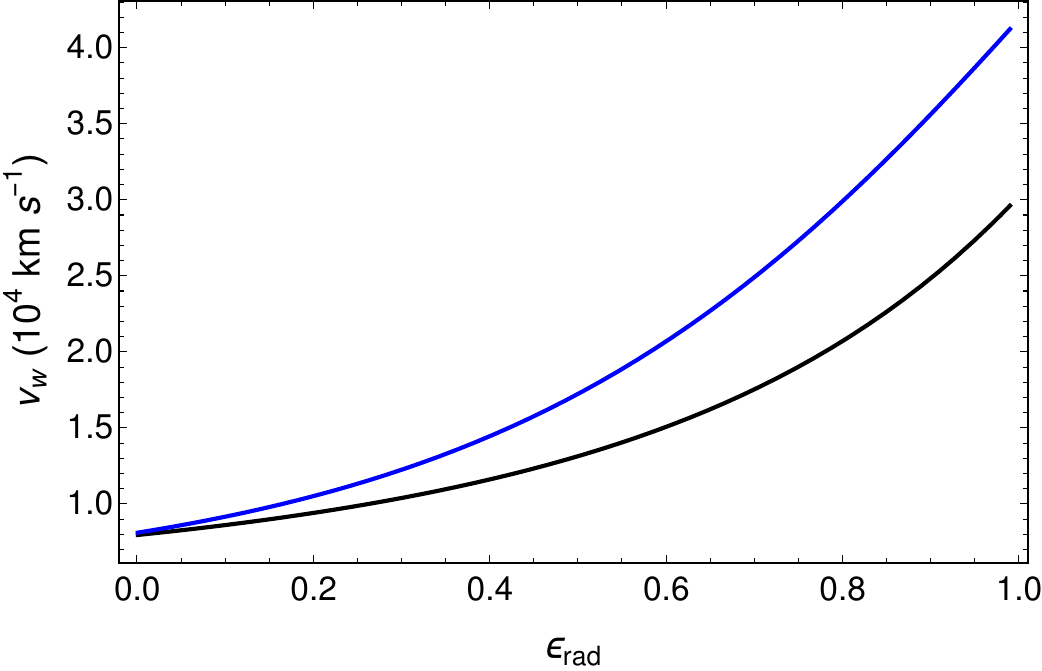}}
\subfigure[]{\includegraphics[scale = 0.44]{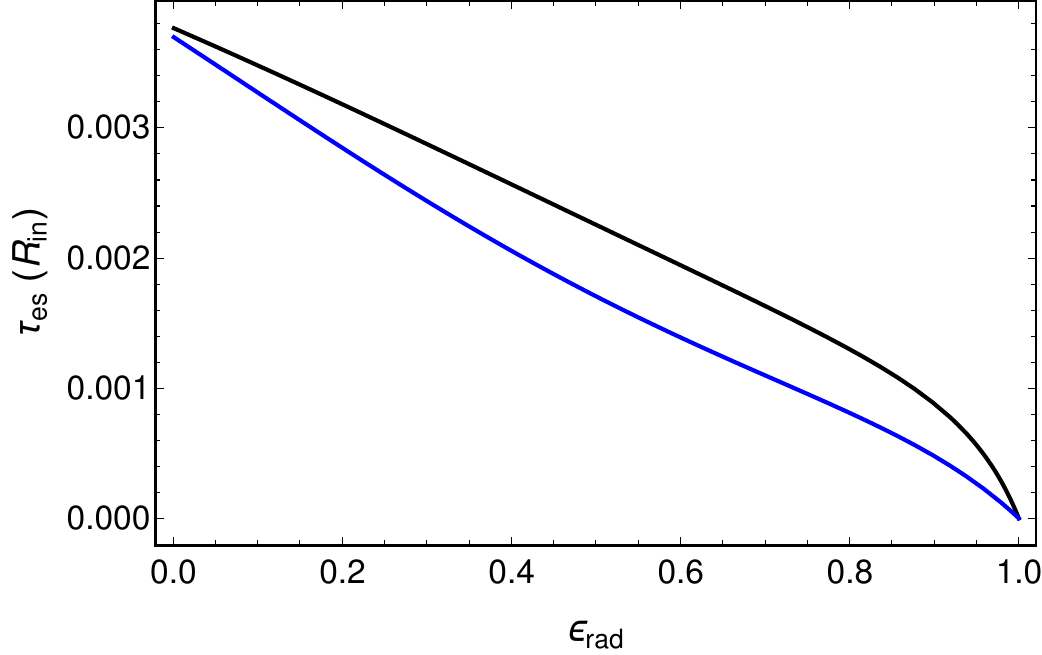}}
\caption{
The same format as Figure \ref{fig:diskwind}, but the dependence is shown on $\epsilon_{\rm rad}$ for two values of $\psi$. 
}
\label{fig:diskwinderad} 
\end{figure}

Figure \ref{fig:diskwind} illustrates the dependence of the total mass loss rate, the wind kinetic energy rate, and the wind velocity on $\psi$ for an AGN disk with $M = 10^7 M_{\odot}$, $\dot{M}_{\rm out} = L_{\rm Edd}/c^2$, and $r_{\rm out} = 10^4 r_{\rm g}$. The figure shows that the total mass loss rate decreases with increasing $\epsilon_{\rm rad}$ and approaches $\dot{M}_{\rm out}$ as $\epsilon_{\rm rad}$ approaches zero, and the wind velocity is of the order of a few $\times 10^4~{\rm km~s^{-1}}$. 
Figure \ref{fig:diskwinderad} shows the dependence of the physical quantities of the wind on $\epsilon_{\rm rad}$ for two values of $\psi$. In the case of no wind ($\epsilon_{\rm rad} = 1$), both the mass loss rate and the wind power take zero values. As $\epsilon_{\rm rad}$ decreases, the mass loss rate increases, as shown in panel (a), leading to an increase in $\dot{E}_{\rm kin}$. In contrast, the mass accretion rate decreases as $\epsilon_{\rm rad}$ decreases, as shown in Figure~\ref{fig:mdotst}. As a result, the increase in the mass loss rate caused by the decrease in $\epsilon_{\rm rad}$ leads to a corresponding decrease in the mass accretion rate, which, in turn, reduces the viscous heating rate of the disk.
Since the viscous heating, in addition to the heating due to magnetic braking, is the source of the wind energy (see equations \ref{eq:qvisf}--\ref{eq:qradf}), the decrease in the viscous heating rate leads to the decrease in $\dot{E}_{\rm kin}$. Therefore, the wind power reaches a maximum at the point where the opposing effects of increasing mass loss rate on the wind power cancel each other out, as shown in panel (b). Moreover, the decrease in the viscous heating flux with decreasing $\epsilon_{\rm rad}$ reduces the radiative flux (see equation \ref{eq:qradf}). This indicates that, in the presence of wind (i.e., as $\epsilon_{\rm rad}$ approaches zero), the disk luminosity decreases regardless of whether $\dot{E}_{\rm kin}$ increases or decreases. Panel (c) depicts the dependence of the wind velocity on $\epsilon_{\rm rad}$. Note that the wind velocity is defined within the range $0 < \epsilon_{\rm rad} < 1$ according to equation~(\ref{eq:vw}).

%
%


%
%

%
%

Assuming that the wind is launched spherically from the disk, the radial density profile of the wind is given by \cite{2009MNRAS.400.2070S,2020ApJ...894....2P,2023MNRAS.518.5693M,2023arXiv231217417M}
\begin{equation}
\rho(R) = \frac{\dot{M}_{\rm w}(r_{\rm out})}{4 \pi R^2 v_{\rm w}}, 
\label{wden}
\end{equation}
where $R$ denotes the spherical radial coordinate, and $\dot{M}_{\rm w}(r_{\rm out})$ and $v_{\rm w}$ are given by equations (\ref{eq:mwtot}) and (\ref{eq:vw}), respectively. The optical depth of the wind with an opacity dominated by electron scattering is given by
\begin{eqnarray}
\tau_{\rm es}(R)
&=& 
\int_{R}^{R_{\rm out}} \kappa_{\rm es} \rho(R) \, \diff R 
\approx
\kappa_{\rm es} \frac{\dot{M}_{\rm w}(r_{\rm out}) }{ 4 \pi v_{\rm w} R}
= 
\sqrt{\frac{2\zeta_2(\zeta_1 -1)}{(1-\epsilon_{\rm rad})(3+\psi)}} \frac{\kappa_{\rm es}\dot{M}_{\rm out}}{4\pi R_{\rm in} \Omega_{\rm out} r_{\rm out}} 
\nonumber \\
&\times&
\left[
1+\frac{\zeta_1}{\zeta_2}\left(\frac{r_{\rm in}}{r_{\rm out}}
\right)^{\delta/2}\right]^{-1} \left[
1+\frac{\zeta_1}{\zeta_2}\left(\frac{r_{\rm in}}{r_{\rm out}}\right)^{\delta/2} - \frac{\delta}{2\zeta_2}\left(\frac{r_{\rm in}}{r_{\rm out}}\right)^{\zeta_1} \right]^{3/2} 
\nonumber \\
&\times& 
\left[1 + \frac{\zeta_1-1}{\zeta_2+1}\left(\frac{r_{\rm in}}{r_{\rm out}}\right)^{\delta/2}-\frac{\delta}{2(\zeta_2+1)}\left(\frac{r_{\rm in}}{r_{\rm out}}\right)^{\zeta_1-1}\right]^{-1/2} \left(\frac{R}{R_{\rm in}}\right)^{-1},
\label{eq:taues}
\end{eqnarray}
where $R_{\rm in}$ and $R_{\rm out}$ are the wind inner and outer radii, respectively, and we assume that $R_{\rm out} \gg R_{\rm in}$. Panel~(d) of Figures~\ref{fig:diskwind} and \ref{fig:diskwinderad} shows the dependence of the electron-scattering-dominated optical depth on $\psi$ and $\epsilon_{\rm rad}$, respectively. Here, $\tau_{\rm es}$ is estimated at $R_{\rm in}$, which is assumed to be equal to the disk's outer radius, $r_{\rm out}$. From the figure, it is evident that the optical depth decreases with increasing $\epsilon_{\rm rad}$, owing to the reduction in the mass loss rate. The wind is highly optically thin, as $\tau_{\rm es}$ is significantly less than unity, implying that photons emitted from the accretion disk pass through the wind media without significant reprocessing.

%
\section{Discussions}
\label{sec:dis}

In this section, we explore key aspects of the one-dimensional, steady-state, magnetically driven accretion disk wind model, focusing on the three constant parameters embedded in the steady-state solutions. We also discuss the dimensionless mass flux parameter, which characterizes the mass outflow in the disk wind due to the multi-dimensionality effect. Finally, we propose a methodology for comparing the theoretically predicted disk spectrum with observational data, a crucial step in validating the accuracy of the steady-state solutions. This approach will test our model and provide a deeper understanding of the physical processes at work in accretion disk winds and their observational signatures.

In our model, the parameters $\epsilon_{\rm rad}$, $\bar{\alpha}_{r\phi}$, and $\psi$ are assumed to be constant. The parameter $\psi$ regulates the vertical stress, $\bar{\alpha}_{z\phi}$, while $\bar{\alpha}_{r\phi}$ represents the Shakura-Sunyaev viscosity parameter, where $\alpha = 2 \bar{\alpha}_{r\phi}/3$ in purely hydrodynamic cases. We adopt $\bar{\alpha}_{r\phi} = 0.1$, which corresponds to $\alpha \approx 0.067$. Observations by Starling et al. (2004) \cite{2004MNRAS.347...67S} suggest that $0.01 \leq \alpha \leq 0.03$, based on the optical variability of quasars over timescales of months to years. This range overlaps with the $\alpha$ values ($\sim 0.005$ to $\sim 0.6$) predicted by MHD turbulent disk simulations, as reported in several studies \cite{1995ApJ...440..742H,1995ApJ...445..767M,1995ApJ...446..741B,2018ApJ...866..134G}. The parameters $\epsilon_{\rm rad}$ and $\psi$ influence disk properties such as surface density, temperature, and mass accretion rate. However, in an ideal MHD disk, these parameters are expected to evolve with the magnetic field rather than remain constant. Therefore, multi-dimensional MHD simulations of disk-wind systems are necessary to accurately determine the behavior of these parameters

In local MHD shearing box simulations of protoplanetary disks, Suzuki \& Inutsuka (2009)~\cite{2009ApJ...691L..49S} showed that magnetorotational instability (MRI) amplifies magnetic fields, leading to the formation of large-scale channel flows where magnetic pressure becomes comparable to gas pressure. The disruption of these flows due to magnetic reconnection drives the disk wind. Suzuki et al. (2010)~\cite{2010ApJ...718.1289S} introduced the dimensionless mass flux $C_{\rm w} = \dot{\Sigma}_{\rm w}/(\rho\,c_{s})_{\rm mid}$ to characterize disk winds, where $\dot{\Sigma}_{\rm w}$ is the mass flux, and $(\rho\,c_{s})_{\rm mid}$ represents the density flux evaluated at the midplane of the disk. The mass flux is controlled by the density at the wind's onset region, located where magnetic energy becomes comparable to thermal energy. Depending on the vertical magnetic field strength, $C_{\rm w}$ typically ranges between $10^{-5}$ and $10^{-4}$. In the context of TDEs, Mageshwaran et al. (2024)~\cite{2024ApJ...975...94T} constructed a one-dimensional, time-dependent, magnetically driven disk wind model. They introduced the dimensionless mass flux, $C_{\rm w,0}$, from energy conservation equation and constrained $C_{\rm w}$ by taking $C_{\rm w} = {\rm Min}[C_{\rm w,0}, C_{\rm w,\rm sim}]$, where $C_{\rm w,\rm sim} = 2\times 10^{-5}$ was estimated from the MHD shearing box simulations noted above.

The dimensionless mass flux can be rewritten using the hydrostatic equilibrium as $C_{\rm w,0} = 2\dot{\Sigma}_{\rm w} / \Sigma \Omega$, where $\dot{\Sigma}_{\rm w}$ and $\Sigma$ are given by equations (\ref{eq:sigdotw2}) and (\ref{eq:sigmast}), respectively, for the steady-state solutions. For $M = 10^7 M_{\odot}$, $\dot{M}_{\rm out} = L_{\rm Edd}/c^2$, $r_{\rm out} = 10^{4} r_{\rm g}$, $\psi = 10$, and $\epsilon_{\rm rad} = 0.1$, we find that $C_{\rm w,0} < 7.13 \times 10^{-6}$, which is smaller than the values estimated by simulations. The value of $C_{\rm w,0}$ is lower at smaller radii and decreases as $\epsilon_{\rm rad}$ increases and $\psi$ decreases. The relationship $C_{\rm w,0} \propto \dot{M}_{\rm out}^{2/5}\Omega_{\rm out}^{3/5}/(r^2 \Omega^2) \propto M^{-1/5}$ holds, where we used equations (\ref{eq:sigdotw2}), (\ref{eq:cssigma2}) and (\ref{eq:sigmast}), noting that $r \propto M$ and $\dot{M}_{\rm out} = L_{\rm Edd} / c^2 \propto M$. This indicates that $C_{\rm w,0}$ decreases with black hole mass. For $M = 10 M_{\odot}$, with other model parameters identical to those used for $M = 10^7 M_{\odot}$, we find $C_{\rm w,0} < 1.14 \times 10^{-4}$. This suggests that $C_{\rm w,0}$ is better constrained within simulation values for higher mass black holes. It is important to note that the simulation was conducted for a protoplanetary disk, and MHD simulations for different black hole masses can provide better constraints on $C_{\rm w}$.

A disk spectrum consists of three frequency regimes: the far-infrared waveband is predominantly in the Rayleigh-Jeans regime, the mid-frequency regime arises from the multicolor component of the blackbody radiation, and the soft X-ray waveband decays exponentially according to Wien's law. In the standard disk model, the spectrum in the multicolor regime follows a power law in frequency, $\nu^{1/3}$, whereas our steady-state solutions yield a spectrum proportional to $\nu^{(1-3\zeta_1)/(3-\zeta_1)}$, where $\zeta_1$ depends on $\epsilon_{\rm rad}$ and $\psi$. As a result, the power-law index becomes lower than $1/3$, and it can even be negative for $\epsilon_{\rm rad}<22/27$ with $\psi=0$ and for $\epsilon_{\rm rad}<2/3$ with $\psi\gg1$. Furthermore, for the negative-index case, the peak frequency lies in the infrared for supermassive black holes and in the near ultraviolet (UV) for stellar-mass black holes at lower $\epsilon_{\rm rad}$, as suggested by the effective temperature in Wien's displacement law. These emissions arising from the outer disk region are unlikely to constrain black hole mass and spin in the presence of disk winds. To test these spectral properties (see also Figure~\ref{fig:sl}), we propose multi-wavelength observations from the near-infrared (NIR) to the optical to the UV with, for instance, Subaru, JWST, and Swift UVOT. Since these facilities detect the spectral fluxes in different wavebands, their data can reveal the spectral slopes over the frequency range, particularly in the multicolor blackbody region. A clear deviation of the power-law index from $1/3$, especially if it is negative, would strongly indicate the presence of a magnetically driven wind from the disk, enabling a test of our disk wind model.

Incidentally, it is known that the disk spectrum in the slim disk model can also have a negative power law index, i.e., $\nu^{-1}$ in the multicolor regime (see Kato et al. 2008 \cite{2008bhad.book.....K} for a review). So how can our model be distinguished from the slim disk model? There are three critical differences between our model and the slim disk models that can help to distinguish them observationally: First, while the slim disk emission has a peak in the soft X-ray bands almost independently of the black hole mass \cite{2006ApJ...648..523W}, the peak of the high-energy side of the emission in our model shifts to the IR--UV region depending on the black hole mass. Next, in our formulation, the spectral slope is $-1$ for the $\epsilon_{\rm rad} = 0$ case, indicating no disk emission and thus being unphysical. For a wide range of $\epsilon_{\rm rad}$, the slope is negative and flatter than $-1$, as shown in Figure~\ref{fig:ple}. Finally, in a radiation-pressure dominated slim disk, the slope of $-1$ appears only when the mass accretion rate is substantially higher than the Eddington rate; otherwise, the slope is not equal to $-1$ because the disk surface temperature does not simply follow $r^{-1/2}$ \cite{2023PhRvD.108d3021M}. Furthermore, slim disk models can exceed the Eddington luminosity, while our model is clearly at a sub-Eddington level. These differences in both spectral slopes and luminosity regimes provide a clear way to distinguish the two scenarios through targeted observations.

\section{Conclusions}
\label{sec:con}
%
We have derived steady-state solutions for a one-dimensional, magnetically-driven accretion disk wind model, constructed based on MHD equations, under the assumption of a geometrically thin, gas-pressure-dominated accretion disk. An extended alpha-viscosity prescription is applied for the turbulent viscosity and magnetic braking. Our model is characterized by three parameters: $\bar{\alpha}_{r\phi}$ for the turbulent viscosity, $\psi$ for magnetic braking, and $\epsilon_{\rm rad}$, which represents the ratio of radiative cooling to disk heating fluxes. Our key results are as follows: 
\begin{enumerate} 
\item 
We confirm that the steady-state solutions for the disk wind reduce to the standard disk solutions when the wind is absent. 
\item 
The mass accretion rate remains constant in the absence of the wind. In contrast, when a wind is present, the mass accretion rate increases with radius, eventually reaching a constant value at the outer boundary, following a power-law dependence on radius.
\item 
The spectral luminosity of the disk peaks in the absence of wind ($\epsilon_{\rm rad} = 1$), while it decreases when a wind is present ($\epsilon_{\rm rad} < 1$). This is consistent with the fact that the mass accretion rate decreases as mass is lost inward in the presence of the wind.
\item 
In the intermediate frequency range, which corresponds to the multicolor part of the disk spectrum, the spectral luminosity follows a power-law relation with frequency as $\propto \nu^{(1-3\zeta_1)/(3-\zeta_1)}$, where $\zeta_1=(1+\psi)[\sqrt{1+8(3+\psi)(1-\epsilon_{\rm rad})/(1+\psi)^2}-1]/4$. The power-law index reduces to $1/3$, which is consistent with the standard disk model when the wind is absent. Notably, the power-law index becomes negative if $\epsilon_{\rm rad}<22/27$ for $\psi = 0$ and $\epsilon_{\rm rad}<2/3$ for $\psi \gg 1$. A negative power-law index strongly supports the presence of a steady-state magnetically driven disk wind.
\item
The mass loss rate increases with radius. The lost mass forms the wind, whose medium is sufficiently optically thin that it does not significantly affect the disk emission.
\end{enumerate}

\section*{Acknowledgment}
We thank the referee for the constructive suggestions that have improved the paper.
M.T. and K.H. have been supported by the Basic Science Research Program through the National Research Foundation of Korea (NRF) funded by the Ministry of Education (2016R1A5A1013277 to K.H. and 2020R1A2C1007219 to K.H. and M.T.). This work is also supported by Grant-inAid for Scientific Research from the MEXT/JSPS of Japan, 22H01263 to T.K.S. This research was supported in part by grant no. NSF PHY-2309135 to the Kavli Institute for Theoretical Physics (KITP).


%

\vspace{0.2cm}
\noindent

\let\doi\relax
\bibliographystyle{ptephy}
\bibliography{reference}

\appendix

%
\section{{Derivation of Basic Equations for MHD Disk and Wind Evolution}}
%

In this appendix, we describe the detailed derivation of the evolutionary equations for the mass and energy of a disk with wind (see also Suzuki et al. 2016~\cite{2016A&A...596A..74S}; Mageshwaran et al. 2024~\cite{2024ApJ...975...94T}).

%
\subsection{{Surface Density Evolution}}
\label{app_sdeq}
%

The mass conservation and momentum conservation equations of the general magneto-hydrodynamics (MHD) 
are given by \cite{1998RvMP...70....1B} as
\begin{equation}\label{mcons}
\frac{\partial \rho}{\partial t} + \vec{\nabla}\cdot (\rho \vec{v}) = 0
\end{equation}
and
\begin{equation}\label{momcons}
\rho \frac{\partial \vec{v}}{\partial t} + (\rho \vec{v}\cdot \vec{\nabla}) \vec{v} = - \vec{\nabla}\left(p + \frac{B^2}{8 \pi}\right) - \rho \vec{\nabla}\Phi + \left(\frac{\vec{B}}{4\pi} \cdot \vec{\nabla}\right) \vec{B} +  \eta_v \left(\nabla^2 \vec{v} + \frac{1}{3} \vec{\nabla}(\vec{\nabla}\cdot \vec{v})\right),
\end{equation}
respectively, where $\vec{v}$ is the fluid velocity, $p$ is the pressure, 
\begin{eqnarray}
\Phi =-\frac{GM}{\sqrt{r^2+z^2}}
\label{eq:gp}
\end{eqnarray}
is the gravitational potential, $B$ is the magnetic-field vector, and $\eta_v$ is the microscopic kinematic shear viscosity. 
Balbus \& Hawley (1998)~\cite{1998RvMP...70....1B} assumed that the bulk viscosity due to the microscopic kinematic shear viscosity $\eta_v$ vanishes. 

Assuming that the disk and wind are axisymmetric, we rewrite equations~\ref{mcons} and \ref{momcons} with cylindrical coordinates as
\begin{equation}\label{cylmcons}
\frac{\partial \rho}{\partial t} + \frac{1}{r} \frac{\partial }{\partial r}(r \rho v_r) + \frac{\partial}{\partial z} (\rho v_z) = 0,
\end{equation}
and
\begin{equation}\label{cylang1}
\frac{\partial}{\partial t}(r \rho v_{\phi}) + \frac{1}{r} \frac{\partial}{\partial r}\left[r^2 \left\{\rho v_r v_{\phi} - \frac{B_r B_{\phi}}{4\pi}\right\}\right] + \frac{\partial}{\partial z}\left[r\left\{\rho v_z v_{\phi} - \frac{B_z B_{\phi}}{4\pi}\right\}\right] = 0,
\end{equation}
{respectively.}

For the purpose of adopting the $\alpha$ prescription \cite{1973A&A....24..337S} for our model, we decompose the azimuthal velocity, $v_{\phi}$, into the mean Keplerian flow and perturbation components as 
\begin{eqnarray}
v_{\phi} = r \Omega + \delta v_{\phi},
\label{eq:vphi}
\end{eqnarray} 
where
$\delta v_{\phi} \ll r \Omega$ {and $\Omega$ is the Keplerian frequency:
\begin{eqnarray}
\Omega=\sqrt{\frac{GM}{r^3}}.
\label{eq:okep}
\end{eqnarray} 
Note that $\Phi=-r^2\Omega^2$ at the disk mid-plane ($z=0$) from equation~(\ref{eq:gp}). The angular momentum conservation equation (\ref{cylang1}) using equation (\ref{eq:vphi}) is given by
\begin{multline}\label{cylang2}
\frac{\partial}{\partial t}(r^2\Omega \rho) + \frac{1}{r}\frac{\partial}{\partial r}\left[r^3\Omega v_r \rho + r^2 \rho \left\{v_{r}\delta v_{\phi}-\frac{B_rB_{\phi}}{4\pi\rho}\right\}\right] + r^2\Omega \frac{\partial}{\partial z}(\rho v_z) \\+ r \frac{\partial}{\partial z}\left[\rho\left\{v_{z}\delta v_{\phi}-\frac{B_zB_{\phi}}{4\pi\rho}\right\}\right] = 0.
\end{multline}  
In addition, we vertically integrate equations (\ref{cylmcons}) and (\ref{cylang2}) to get 
\begin{equation}\label{cylmcons1}
\frac{\partial \Sigma}{\partial t} + \frac{1}{r}\frac{\partial}{\partial r}(r \Sigma v_r) + \dot{\Sigma}_{\rm w}  = 0
\end{equation}
and
\begin{multline}\label{cylang3}
\frac{\partial}{\partial t}(r^2\Omega \Sigma) + \frac{1}{r} \frac{\partial}{\partial r} \left[r^3 \Omega v_r \Sigma + r^2 \int \rho\left\{v_{r}\delta v_{\phi}-\frac{B_rB_{\phi}}{4\pi\rho}\right\}\, \diff z \right] + r^2 \Omega \dot{\Sigma}_{\rm w} \\ + r \int \frac{\partial}{\partial z}\left[\rho\left\{v_{z}\delta v_{\phi}-\frac{B_zB_{\phi}}{4\pi\rho}\right\}\right] \, \diff z = 0,
\end{multline}
where 
\begin{equation}\label{eq:sig2Hrho}
\Sigma = 2 H \rho
\end{equation}
is the surface density in the disk, 
\begin{eqnarray}\label{eq:vmflux}
\dot{\Sigma}_{\rm w} = 2\rho v_{z,H}
\end{eqnarray}
is the vertical mass flux, and $v_{z,H} \equiv v_{z}(r,H)$ is the vertical velocity evaluated at the disk scale-height $H$. Using $\alpha$ prescription and following Suzuki et al. (2016)~\cite{2016A&A...596A..74S}, we take
\begin{eqnarray}
\bar\alpha_{r\phi}
&\equiv&
\frac{1}{c_{s}^2}
\int_{-H}^{H} \rho \left[v_r \delta v_{\phi} - \frac{B_r B_{\phi}}{4\pi \rho}\right] 
\, dz 
\biggr/
\int_{-H}^{H} \rho\, dz, \label{eq:alrphi0} \\
\nonumber \\
\bar\alpha_{z\phi} 
&=&
\frac{1}{c_{s}^2}
\biggr[
v_z \delta v_{\phi} - \frac{B_z B_{\phi}}{4\pi \rho}
\biggr]_{z=-H}^{z=H},
\label{eq:alzphi1} 
\end{eqnarray}
where $c_s^2$ is the sound speed at the mid-plane and $\bar\alpha_{r\phi}$ and $\bar\alpha_{z\phi}$ control the angular momentum flux transport radially and vertically in the disk. Using these $\alpha$ parameters, equation (\ref{cylang3}) is given by
\begin{equation}\label{cylang4}
\frac{\partial}{\partial t}(r^2\Omega \Sigma) + \frac{1}{r} \frac{\partial}{\partial r} \left[r^2 \Sigma\left\{v_r r \Omega + \bar{\alpha}_{r\phi} c_s^2 \right\}\right] + r \left[\dot{\Sigma}_{\rm w}  r \Omega + \bar{\alpha}_{z\phi} \rho c_s^2\right] = 0,
\end{equation}

With Keplerian angular rotational velocity given by equation (\ref{eq:okep}), equation (\ref{cylang4}) is reduced to
\begin{equation}\label{rsvreqn}
r \Sigma v_r = -\frac{2}{r\Omega} \left[\frac{\partial}{\partial r} \left(r^2 \Sigma \bar{\alpha}_{r\phi} c_s^2 \right) + r^2 \bar{\alpha}_{z\phi} \rho c_s^2\right],
\end{equation} 
where equations~(\ref{eq:okep}) and (\ref{cylmcons1}) are used for the derivation. Equation~(\ref{rsvreqn}) makes it possible to evaluate the mass accretion rate of the disk: $\dot{M} = -2 \pi r \Sigma v_r$. Substituting equation (\ref{rsvreqn}) into equation (\ref{cylmcons1}), we obtain the evolutionary equation of the surface density as
\begin{equation}\label{sigteqn}
\frac{\partial \Sigma}{\partial t} - \frac{2}{r}\frac{\partial}{\partial r}\left[\frac{1}{r\Omega}\left\{\frac{\partial}{\partial r} \left(r^2 \Sigma \bar{\alpha}_{r\phi} c_s^2 \right) + r^2 \bar{\alpha}_{z\phi} \rho c_s^2 \right\}\right] + \dot{\Sigma}_{\rm w} = 0,
\end{equation}

%
\subsection{{Energy Equation}}
\label{app_eneq}
%

The energy conservation equation is given by \cite{1998RvMP...70....1B}

\begin{equation}
\frac{\partial}{\partial t}\left[\frac{1}{2} \rho v^2 + \rho \Phi + \frac{p}{\gamma -1} + \frac{B^2}{8\pi}\right] + \vec{\nabla} \cdot \left[\vec{v} \left(\frac{1}{2} \rho v^2 + \rho \Phi + \frac{\gamma}{\gamma -1}p \right)  + \frac{\vec{B}}{4\pi} \times \left(\vec{v} \times \vec{B}\right) + \vec{Q} \right] = 0,
\nonumber
\end{equation}

\noindent where $\gamma$ is a ratio of specific heats and {$\vec{Q}=(Q_r,Q_\phi,Q_z)$} is other contributions to energy flux in addition to the MHD energy, such as thermal conduction and radiative heating or cooling. 
{The above equation is rewritten in cylindrical coordinates with the axisymmetric assumption as}
\begin{multline}\label{eneqn}
\frac{\partial}{\partial t}\left[\frac{1}{2} \rho v^2 + \rho \Phi + \frac{p}{\gamma -1} + \frac{B^2}{8\pi}\right] + \frac{1}{r} \frac{\partial}{\partial r}\left[r \left\{v_r \left(\frac{1}{2} \rho v^2 + \rho \Phi + \frac{\gamma}{\gamma -1}p + \frac{B_{\phi}^2 + B_z^2}{4\pi}\right) \right.\right. \\ \left.\left.- \frac{B_r}{4\pi} \left(v_{\phi}B_{\phi} + v_z B_z\right) + Q_{r}\right\}\right] + \frac{\partial}{\partial z}\left[v_z \left(\frac{1}{2} \rho v^2 + \rho \Phi + \frac{\gamma}{\gamma -1}p + \frac{B_{\phi}^2 + B_r^2}{4\pi}\right) \right. \\ \left.- \frac{B_z}{4\pi} \left(v_{\phi}B_{\phi} + v_r B_r\right) + Q_{z}\right] = 0.
\end{multline}

{
Assuming $r \Omega \gg v_r,~\delta v_{\phi},~v_z,~c_s,~B/\sqrt{4\pi \rho}$, the second and third terms of equation (\ref{eneqn}) are reduced to
} 
\begin{multline}\label{mnt3}
\frac{\partial}{\partial r}\left[r \left\{v_r \left(\frac{1}{2} \rho v^2 + \rho \Phi + \frac{\gamma}{\gamma -1}p + \frac{B_{\phi}^2 + B_z^2}{4\pi}\right) - \frac{B_r}{4\pi} \left(v_{\phi}B_{\phi} +  v_z B_z\right)\right\}\right] \\ =\frac{\partial }{\partial r}\left[r \left\{- \frac{1}{2} \rho r^2 \Omega^2 v_r +  \rho r \Omega \left(v_r \delta v_{\phi} - \frac{B_r B_{\phi}}{4\pi \rho}\right) \right\}\right]
\end{multline}
and
\begin{equation}\label{mnt4}
\frac{\partial}{\partial z}\left[v_z \left(\frac{1}{2} \rho v^2 + \rho \Phi + \frac{\gamma}{\gamma -1}p + \frac{B_{\phi}^2 + B_r^2}{4\pi}\right) - \frac{B_z}{4\pi} \left(v_{\phi}B_{\phi} + v_r B_r\right) \right] = \frac{\partial}{\partial z}(\rho v_z E_{\rm w}),
\end{equation}
{
respectively, where equations~(\ref{eq:gp}), (\ref{eq:vphi}), and (\ref{eq:okep}) are adopted for these modifications, 
and $E_{\rm w}$ is given as the wind energy by
}
\begin{equation}\label{eq:ew0}
E_{\rm w} = \frac{1}{2} v^2 + \Phi + \frac{\gamma c_s^2}{\gamma -1} + \frac{B_{\phi}^2 + B_r^2}{4\pi \rho} - \frac{B_z}{4\pi\rho v_z} \left(v_{\phi}B_{\phi} + v_r B_r\right).
\end{equation}

{Substituting} equations (\ref{mnt3}) and (\ref{mnt4}) {into} equation (\ref{eneqn}), we get
\begin{equation}\label{eq:ene}
\frac{\partial}{\partial t}\left[-\frac{1}{2} \rho r^2 \Omega^2\right] + \frac{1}{r}\frac{\partial }{\partial r}\left[r \left\{- \frac{1}{2} \rho r^2 \Omega^2 v_r + \rho r \Omega 
\left(
v_r \delta v_{\phi} - \frac{B_r B_{\phi}}{4\pi \rho}
\right) 
\right\} \right] + \frac{\partial}{\partial z}(\rho v_z E_{\rm w} + Q_{z}) = 0,
\end{equation}
where $Q_r=0$ is adopted because there is no energy dissipation in the radial direction. Integrating equation~(\ref{eq:ene}) vertically leads to
\begin{equation}\label{eq:ene-intz}
\frac{\partial}{\partial t}\left[-\frac{1}{2} \Sigma r^2 \Omega^2\right] + \frac{1}{r}\frac{\partial }{\partial r}\left[r \left\{- \frac{1}{2} \Sigma r^2 \Omega^2 v_r + \Sigma r \Omega \bar{\alpha}_{r\phi} c_s^2 \right\}\right]  + \dot{\Sigma}_{\rm w} E_{\rm w} + {Q}_{\rm rad} = 0,
\end{equation}
where {$Q_{\rm rad} = \int_{-H}^{H}\,Q_{z}\,dz$ is the radiative cooling rate.
%
%
Combining equation (\ref{eq:ene-intz}) with equation~(\ref{sigteqn}) results in 
\begin{equation}
\dot{\Sigma}_{\rm w} \left[E_{\rm w} + \frac{r^2 \Omega^2}{2} \right] + {Q}_{\rm rad} = \frac{3}{2}\Omega \Sigma \bar{\alpha}_{r\phi} c_s^2 + \bar{\alpha}_{z\phi} r \Omega \rho c_s^2,
\end{equation}
where equation~(\ref{eq:okep}) is adopted for the derivation.
}

%
\section{
Rosseland Mean Opacities
}
\label{app:opacity}
%

In Section~\ref{sec:steadystate}, we have derived the steady state solutions for Thomson opacity which is due to the scattering of electromagnetic radiation by free, non-relativistic electrons. Here we present the solution for the following opacity formula: 
\begin{equation}\label{eq:opacity}
\kappa=\kappa_0 \rho^{a} T^{b},
\end{equation}
where $\kappa_0$, $a$ and $b$ are constants (e.g., see Cannizzo et al.1990 \cite{1990ApJ...351...38C}). 
While $\kappa_0=\kappa_{\rm es}$ with $a=0$ and $b=0$ reduces to the electron scattering opacity, $\kappa_0=\kappa_{\rm ff}$ or $\kappa_0=\kappa_{\rm bf}$ with $a=1$ and $b=-7/2$ represents the Rosseland mean opacities for the free-free absorption or bound-free absorption, respectively. Substituting equation~(\ref{eq:opacity}) into equation~(\ref{eq:qrad}) yields the radiative cooling rate as
\begin{equation}\label{eq:qradkappa}
Q_{\rm rad} = \frac{64\sigma T^{4}}{3 \kappa \Sigma} 
= 
\frac{2^{a+6}}{3} 
\frac{\sigma}{\kappa_0} 
\left(\frac{\mu m_{\rm p}}{k_{\rm B}}\right)^{4-b} 
\frac{c_{\rm s}^{a-2b+8}}{\Sigma^{a+1} \Omega^a},
\end{equation}
where we use equation (\ref{eq:cs}) and $\Sigma = 2H\rho$ for the derivation. Equating equation~(\ref{eq:qradkappa}) to equation~(\ref{eq:qradf}) with equation~(\ref{eq:qvisf2}) gives the relation between the sound speed and the surface density as
\begin{equation}
\label{eq:cs-sigma}
c_{\rm s}^{a-2b+6} 
=
\frac{9}{2^{7+a}} 
\frac{\kappa_0}{\sigma}
\left(\frac{k_{\rm B}}{\mu m_{\rm p}}\right)^{4-b} 
\bar{\alpha}_{r\phi} 
\epsilon_{\rm rad}
\left(1+\frac{\psi}{3}\right) \Omega^{1+a} \Sigma^{a+2}
\end{equation}

In our steady-state model, the derivation and assumptions of equations (\ref{eq:maccrate2})-(\ref{eq:cssigma2}) are not influenced by opacity. In other words, the solutions for the mass accretion rate, mass loss rate, and heating fluxes remain unaffected by opacity. However, opacity does influence the radial profiles of the disk surface density and the mid-plane temperature of the disk. Combining equation~(\ref{eq:cs-sigma}) with equations~(\ref{eq:cssigma2}) gives the surface density as
\begin{eqnarray}
\label{eq:sigma-opacity}
\Sigma 
&=&
\pi^{-(a-2b+6)/(3a-2b+10)}
\left(\frac{2^{1+2b}}{3} \right)^{2/(3a-2b+10)}
\left[
\frac{\sigma}{\kappa_0}
\left(\frac{\mu m_{\rm p}}{k_{\rm B}}\right)^{4-b} 
\right]^{2/(3a-2b+10)}
\nonumber \\
&\times&
\biggr[
\bar{\alpha}_{r\phi}^{(a-2b+8)}
\zeta_2^{a-2b+6}
\bigr[\epsilon_{\rm rad}(3+\psi)\bigr]^2
 \biggr]^{-1/(3a-2b+10)}
\dot{M}_{\rm out}^{(a-2b+6)/(3a-2b+10)} 
\Omega_{\rm out}^{-(a+2b-4)/(3a-2b+10)} 
\nonumber \\
&\times&
\left[1+\frac{\zeta_1}{\zeta_2} \left(\frac{r_{\rm in}}{r_{\rm out}}\right)^{\delta/2}\right]^{-(a-2b+6)/(3a-2b+10)} 
\left[1-\left(\frac{r_{\rm in}}{r}\right)^{\delta/2}\right]^{(a-2b+6)/(3a-2b+10)}
\nonumber \\
&\times&
\left(\frac{r}{r_{\rm out}}\right)^{
\bigr[
\zeta_1(a-2b+6)+\frac{3}{2}(a+2b-4)
\bigr]
/
(3a-2b+10)
} .
\end{eqnarray}

Combining equation~(\ref{eq:cs}) with equations~(\ref{eq:cssigma2}) and (\ref{eq:sigma-opacity}) yields the disk mid-plane temperature, $T = (\mu m_p/k_{\rm B}) (c_s^2\Sigma/\Sigma)$, as 
\begin{eqnarray}
T
&=&
\pi^{-2(a+2)/(3a-2b+10)}
2^{-2(3a+11)/(3a-2b+10)}
3^{2/(3a-2b+10)}
\nonumber \\
&\times&
\left(
\frac{\sigma}{\kappa_0}
\right)^{-2/(3a-2b+10)}
\left(
\frac{\mu m_{\rm p}}{k_{\rm B}}\right)^{(3a+2)/(3a-2b+10)}
\nonumber \\
&\times&
\bar{\alpha}_{r\phi}^{-2(a+1)/(3a-2b+10)}
\zeta_2^{-2(a+2)/(3a-2b+10)}
\bigr[\epsilon_{\rm rad}(3+\psi)\bigr]^{2/(3a-2b+10)}
 \nonumber \\
 &\times&
\dot{M}_{\rm out}^{2(a+2)/(3a-2b+10)} 
\Omega_{\rm out}^{2(2a+3)/(3a-2b+10)} 
\nonumber \\
&\times&
\left[1+\frac{\zeta_1}{\zeta_2} \left(\frac{r_{\rm in}}{r_{\rm out}}\right)^{\delta/2}\right]^{-2(a+2)/(3a-2b+10)} 
\left[1-\left(\frac{r_{\rm in}}{r}\right)^{\delta/2}\right]^{2(a+2)/(3a-2b+10)}
\nonumber \\
&\times&
\left(\frac{r}{r_{\rm out}}\right)^{
\left[\zeta_1(2a+4)-(6a+9)\right]/(3a-2b+10)} .
\label{eq:temp-opacity}
\end{eqnarray}

For $\kappa_0=\kappa_{\rm es}$ with $a = 0$ and $b=0$, equations~(\ref{eq:sigma-opacity}) and (\ref{eq:temp-opacity}) reduce to the Thomson opacity solutions, i.e., equations~(\ref{eq:sigmast}) and (\ref{eq:tmidst}), respectively. In the Rosseland mean opacity, expressed using Kramers' law for free-free absorption,
$\kappa_0 = 6.24 \times 10^{24}~{\rm cm^5~K^{7/2}~g^{-2}}$ with $a = 1$ and $b = -7/2$ \cite{2008bhad.book.....K}, the surface density and temperature are given by
\begin{eqnarray}
\Sigma 
&=&
\frac{1}{2^{3/5}(3\pi^7)^{1/10}}
\left(
\frac{\sigma}{\kappa_0}
\right)^{1/10}
\left(
\frac{\mu m_{\rm p}}{k_{\rm B}}
\right)^{3/4} 
\bar{\alpha}_{r\phi}^{-4/5}
\zeta_2^{-7/10}
\bigr[
\epsilon_{\rm rad}
(3+\psi)\bigr]^{-1/10}
\dot{M}_{\rm out}^{7/10} 
\Omega_{\rm out}^{1/2} 
\nonumber \\
&\times&
\left[1+\frac{\zeta_1}{\zeta_2} \left(\frac{r_{\rm in}}{r_{\rm out}}\right)^{\delta/2}\right]^{-7/10} 
\left[1-\left(\frac{r_{\rm in}}{r}\right)^{\delta/2}\right]^{7/10}
\left(\frac{r}{r_{\rm out}}\right)^{7\zeta_1/10-3/4}.
\label{eq:sigma_out}
\end{eqnarray}
and 
\begin{eqnarray}
T
&=&
\frac{1}{2^{7/5}}
\left(
\frac{3}{\pi^3}
\right)^{1/10}
\left(
\frac{\kappa_0}{\sigma}
\right)^{1/10}
\left(
\frac{\mu m_{\rm p}}{k_{\rm B}}\right)^{1/4}
\bar{\alpha}_{r\phi}^{-1/5}
\zeta_2^{-3/10}
\bigr[\epsilon_{\rm rad}(3+\psi)\bigr]^{1/10}
\dot{M}_{\rm out}^{3/10} 
\Omega_{\rm out}^{1/2} 
\nonumber \\
&\times&
\left[1+\frac{\zeta_1}{\zeta_2} \left(\frac{r_{\rm in}}{r_{\rm out}}\right)^{\delta/2}\right]^{-3/10} 
\left[1-\left(\frac{r_{\rm in}}{r}\right)^{\delta/2}\right]^{3/10}
\left(\frac{r}{r_{\rm out}}\right)^{3\zeta_1/10-3/4},
\label{eq:temp_out}
\end{eqnarray}
respectively.
For no wind solution with ($\epsilon_{\rm rad}=1, \psi=0, \zeta_1=0, \zeta_2=1/2, \delta=1$), 
equations~(\ref{eq:sigma_out}) and (\ref{eq:temp_out}) reduce to
\begin{eqnarray}
\Sigma 
&=&
\frac{2^{1/10}}{(9\pi^7)^{1/10}}
\left(
\frac{\sigma}{\kappa_0}
\right)^{1/10}
\left(
\frac{\mu m_{\rm p}}{k_{\rm B}}
\right)^{3/4} 
\bar{\alpha}_{r\phi}^{-4/5}
%
\dot{M}_{\rm out}^{7/10} 
\Omega_{\rm out}^{1/2} 
\left[1-\left(\frac{r_{\rm in}}{r}\right)^{1/2}\right]^{7/10}
\left(\frac{r}{r_{\rm out}}\right)^{-3/4}.
\end{eqnarray}
and 
\begin{eqnarray}
T
&=&
\frac{1}{2^{11/10}}
\left(
\frac{9}{\pi^3}
\right)^{1/10}
\left(
\frac{\kappa_0}{\sigma}
\right)^{1/10}
\left(
\frac{\mu m_{\rm p}}{k_{\rm B}}\right)^{1/4}
\bar{\alpha}_{r\phi}^{-1/5}
\dot{M}_{\rm out}^{3/10} 
\Omega_{\rm out}^{1/2} 
\left[1-\left(\frac{r_{\rm in}}{r}\right)^{1/2}\right]^{3/10}
\left(\frac{r}{r_{\rm out}}\right)^{-3/4},
\end{eqnarray}
respectively. For given $M$, $\dot{M}_{\rm out}$, $r_{\rm in}$, and $r_{\rm out}$, these two solutions correspond to the standard disk solutions in the outer regions where the free-free absorption opacity is dominant
\cite{1973A&A....24..337S,2002apa..book.....F,2008bhad.book.....K}.

\end{document}